\documentclass[twocolumn]{svjour3}          
\smartqed  
\usepackage{setspace}
\usepackage{caption}
\usepackage{amsmath}
\usepackage[Algorithm]{algorithm}
\usepackage{algpseudocode} 
\usepackage{epstopdf}
\usepackage{color}
\usepackage{amssymb}
\usepackage[figuresright]{rotating}
\usepackage{colortbl}
\usepackage{setspace}
\usepackage{caption}
\usepackage{textcomp}
\usepackage{supertabular}
\usepackage{lscape}
\usepackage[figuresright]{rotating}

\begin{document}

\title{A Classification Framework for TDMA Scheduling techniques in  WSNs}

\titlerunning{Two-Phase TDMA-Scheduling}        

\author{Ashutosh Bhatia       \and
        R. C. Hansdah 
}


\institute{Ashutosh Bhatiar \at
               Dept. of CSIS, BITS Pilani \\
               \email{ashutosh.bhatia@pilani.bits-pilani-ac.in} 
               \and
               R. C. Hansdah \at
              Dept. of CSA, IISc Bangalore \\
	   \email{hansdah@csa.iisc.ernet.in}
}

\date{}

\maketitle

\begin{abstract}
One of the major challenges in wireless sensor networks (WSNs) is the mitigation of collisions due to simultaneous transmissions by multiple nodes over a common channel which are located  in a proximity.  TDMA-based channel access provides  energy-efficient and collision-free  transmissions. It  is  especially suitable for traffic with periodic transmission pattern and  guaranteed QoS requirements.  For that reason, a large number of TDMA-scheduling algorithms are available in the literature, and consequently,  a good number of  survey papers on  TDMA-scheduling  algorithms have been written. In this work, we propose  a novel classification framework to categorize the existing TDMA-scheduling algorithms available for WSNs. As against existing survey works,  the proposed framework possess certain new dimensions (categories)  to classify existing TDMA-scheduling algorithms. Additionally, we introduce a couple of  new sub-categories for the existing classes which would help researchers to even differentiate between two TDMA-Scheduling algorithms that are assumed to be similar as per existing classification schemes. Finally,  we also discuss few important works in the context of proposed classification scheme.
\end{abstract}

\section{Introduction}
The TDMA-based media access and control (MAC) protocols are better suited for WSNs against 
collision-avoidance based MAC protocols, especially under peak load
conditions. TDMA-based channel access eliminates collisions, idle listening and overhearing, which are the major cause
of energy consumption in WSNs.  
In addition, it  provides guaranteed end-to-end delay performance in multi-hop wireless
communication. For instance, in WSNs, TDMA-based channel access ensures the timely detection of events at the base station.  Another important aspect of TDMA-based communication is  its superior performance during heavy loads. When the data rate of each node in WSN is high or there are too many sensor nodes, the contention-based MAC protocols  may lead to a large number of  retransmissions and collisions, thereby degrading the energy consumption and the Quality of Service (QoS) performance of the network.

Generating a conflict-free  schedule (TDMA- scheduling) in a  TDMA-based MAC  protocol, is an 
challenging and important problem.  A TDMA-schedule strongly effects the  efficiency of  the channel utilization. In general, the problem of TDMA-scheduling in WSNs can be seen 
as the assignment of time slots to the nodes for their data transmission, ensuring  the transmission of
packets by a node in its assigned time slot, should not collide with  transmission of any other node in the network.
However, TDMA-scheduling for WSNs sometimes considered as 
an  integrated problem with other sub-problems such as finding the path of communication
between two nodes which are at the multi-hop distance from each other (routing problem) and 
efficient utilization of available bandwidth, when more than one communication channels
are available (channel assignment problem). In addition, the TDMA-scheduling problem may have to take into account 
the application constraints, such as  providing quality of service (QoS).

In this paper,  we propose a classification framework for various TDMA-scheduling techniques  based on
certain characteristics which have already been  used in earlier surveys, along with few additional 
characteristics identified by us. The identified characteristics are then categorized further to get a better 
understanding of existing TDMA-scheduling algorithms, using an approach based on 
the objectives which an algorithm try to achieve, the assumptions made by the algorithm and 
design methodology adapted by the algorithm to perform TDMA-scheduling. Thereafter, using the proposed framework, we   give a brief survey of the TDMA-based MAC protocols and algorithms designed for WSNs and few the algorithms which have been designed  for general purpose multi-hop wireless networks instead of WSNs specifically.    

The rest of the paper is organized as follows. First, in section 2, we propose a framework to classify the TDMA-scheduling algorithms 
in WSNs based on the  identified characteristics. In section 2 and 3, we discuss the centralized and distributed existing TDMA-scheduling algorithms respectively in the context of proposed classification framework. Section 4
concludes the paper.

\section{A Framework for the Classification of  TDMA-scheduling Algorithms in WSNs}
\label{section:classification}
TDMA-scheduling  in WSNs  has been the subject of intensive research during 
the last two decades. As a result, a  number of survey papers are available in the literature 
related to TDMA-scheduling in WSNs \cite{survey3-Incel,survey3-Gabale,survey3-Wang}. 
In order to have a classification framework for TDMA-scheduling algorithms in WSNs,
we extend the classification of TDMA-scheduling algorithms for WSNs, provided by  bhaskaran et. el. in 
\cite{survey3-Incel}. The classification presented in \cite{survey3-Incel} is based on scheduling 
objectives and underlying assumptions made by the algorithms. The survey paper covers  scheduling 
algorithms which are designed only for  convergecast communication  in WSNs.  Moreover, the algorithms 
discussed in \cite{survey3-Incel}, are mostly centralized in nature, and generate optimal schedule 
with respect to a given objective such as minimizing energy consumption,  minimizing  latency of data collection  and  minimizing schedule length.

\begin{figure*}[t]
\centering
\includegraphics[scale=.5]{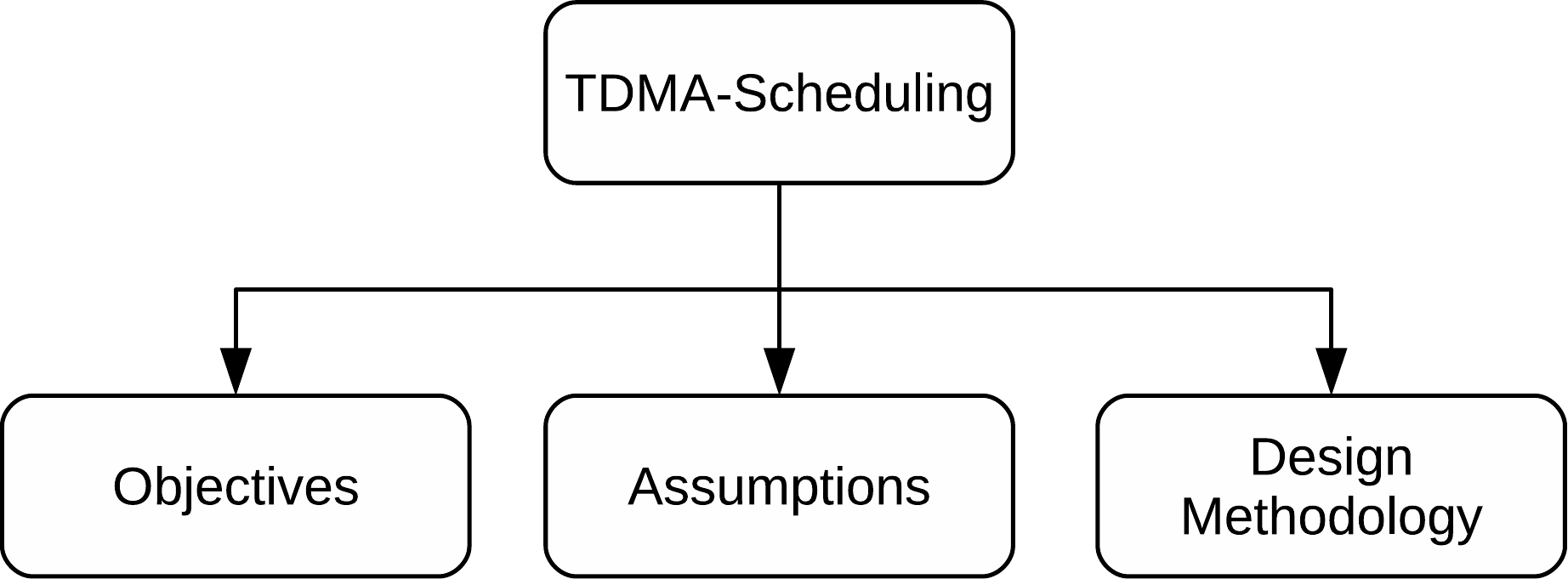}
  \caption{Classification of TDMA-scheduling algorithms}
\label{fig:classification}
\end{figure*}

In addition to the objectives and assumptions, identified as the two dimensions to characterize a TDMA-scheduling algorithm, 
we add ``Design-Methodology'' as the third dimension (Fig. \ref{fig:classification}), to the 
classification presented in \cite{survey3-Incel}. Moreover, we have further classified  the set of  
objectives and assumptions into a multilevel hierarchy, to  better  understand the  design of 
existing TDMA-scheduling algorithms. Finally, we have identified a couple of new objectives and assumptions, 
to cover a larger spectrum of TDMA-scheduling algorithm in WSNs.

In the following subsections, we  discuss the characteristics of each category and sub-category of the proposed
classification framework, in detail.

\subsection{Objectives}
Primary objective of any TDMA-scheduling algorithm is always to find a feasible (or 
interference free) schedule. However, the choice of a  schedule among all feasible schedules 
may differ based on the scheduling objectives set by an algorithm. We further classify the objectives of  TDMA-scheduling 
into two categories: \textit{MAC-performance} objectives and \textit{schedule-characteristics} objectives. The TDMA-scheduling 
objectives belonging to \textit{MAC-performance} category can be seen as the objective to improve the performance of MAC 
protocol with respect to a given parameter (e.g. throughput, energy etc.). In this sense, the objective of a TDMA-scheduling 
algorithm is same as the objective of underlying MAC protocol, in which, the generated schedule is going to be used. 
The objectives belonging to the category \textit{schedule-characteristics} can be seen as the desired property of 
the schedule to be generated by the TDMA-algorithm. For example, a TDMA-scheduling algorithm may aim to generate a 
TDMA-schedule with partially conflict free property to avoid the collisions up to a certain extent, instead of completely 
suppressing the collisions by generating a fully conflict free schedule. Now, we discuss both the categories of 
TDMA-scheduling objectives in detail.

\subsubsection{MAC-performance objectives}
The main objective of a TDMA-scheduling is to find a conflict free schedule, which can improve the 
performance of MAC protocol with respect to given parameters. Therefore, \textit{MAC-performance} objectives of a 
TDMA-scheduling algorithm are same as the overall design objectives of the underlying MAC protocol.  Here, we discuss various 
MAC-performance design objectives that have been considered by various existing TDMA-scheduling algorithms and the implications
of these objectives on the design of  scheduling algorithms.

\begin{itemize}
 \item \textbf{Throughput:}

Improving the throughput performance of a MAC protocol 
is one of the most researched design objective of many TDMA-scheduling  algorithms available in literature. 
The objective of improving  
throughput is mainly achieved by one of the following three techniques.

\begin{itemize}
 \item \textbf{Minimizing Schedule Length:}  In this method, an algorithm tries to minimize the  
schedule length, given the constraint that each node should get exactly one slot per schedule 
duration. Typically, minimizing the scheduling length is equivalent to maximizing the throughput. Additionally, many times, 
a minimal schedule length also implies minimal  end to end latency. 
However, the  number of hops and the sequence of transmissions from source to destination can also be considered together with the scheduling constraints to further reduce  end-to-end latency.

\item \textbf{Maximizing Transmission Set:} In this method, an algorithm tries to maximize the 
number of concurrent transmissions in each slot of a fixed schedule length. Consequently, a  
node may get more than one slot per frame. As compared to previous method, this method does not 
guarantee fairness in the allocation of available bandwidth among the nodes, when all nodes are having equal 
demand. 

\item \textbf{Dynamic Scheduling:} In case of variable load conditions, the TDMA-based channel 
access gives much  higher delays and lower throughput due to  the  static allocation of time slots. A node can use only the time slots allocated to it, even if the  time slots allocated to the other nodes are not being used by them.  In order to improve 
the throughput in such situations, the dynamic scheduling method is applied, in which, the task of  
scheduling is performed  per slot basic or per frame basic (depending upon the current 
transmission requirement of the nodes) instead of using a fixed schedule for a very long time (multiple frames). 
A  dynamic TDMA-scheduling algorithm should take lesser 
time to generate a schedule, otherwise the overhead of scheduling would become very high, as 
the algorithm has to run either in every slot or at the start of every frame.  

\end{itemize}

\item \textbf{Latency:}
It is important to reduce end-to-end latency especially for those applications in which completing a certain task before the deadline is very crucial. Many mission-critical and even-based applications require lesser end-to-end latency as compared to other applications.
Moreover, the algorithms with the goal to minimize the schedule-length may not generate the TDMA-schedules with minimum delay for    few  topologies. 
For example, a linear deployment of sensor nodes  may result  higher spatial  bandwidth utilization  in WSNs.  But,  due to  a large distance (number of hops) between the source to the destination,  packet transmissions may experience higher end-to-end delay. Therefore, to minimizing the end-to-end latency, few additional constraints are  required to be considered,  besides the constraints related to  minimizing the schedule-length.

\item \textbf{Energy Consumption:}
Maximizing the network  lifetime is a crucial requirement for any resource constraint WSNs and this is typically achieved by efficiently  managing the  radio activity of the sensor nodes. 
One of the most common techniques to save the energy of sensor nodes is to perform sleep scheduling  i.e.  periodic switching of sensor radio between sleep and active modes. 
This can be easily achieved 
in TDMA-based MAC protocols, where the nodes have to wake-up only during the time slots in which 
they are either transmitter or the intended receiver of a transmitted packet. 
Transmission power control (TPC)  is also well known technique that is used in   conjunction with
TDMA-scheduling, to control the energy consumption as well as  level of interference in a network. Transmitting packets at maximum available power can cause higher level of inference thereby reducing the capacity of the network. On the other hand, transmitting packets at very low power would  possibly increase the communication delay in the network as size of neighborhood for each node would decrease. Additionally, the quality of wireless links is time dependent, and in this case, dynamic power control can improve the packet delivery rate by  improving the quality of poor links.

\item \textbf{Self Stabilization}: 
In addition to  avoiding the collisions, self stabilization against the changes in the network (such 
as arrival of new nodes), is also an equally important and desired property of a TDMA-based MAC 
protocol. It is not cost effective to perform re-scheduling of  the complete network every time when 
the network topology changes even within a very small portion of the  network. Furthermore, the process of self 
stabilization due to changes at  some part the network should affect only the nodes in the 
vicinity of the change. In literature, capability of self stabilization of a TDMA-scheduling algorithm 
is usually measured in terms of the time taken by the algorithm  to 
reach a conflict free schedule, starting from the time when the change occurred, and the  amount of control 
messages exchanged in the recovery process.

\item \textbf{Communication Overhead:}
In many TDMA-sche-duling algorithms, the process of message exchange with neighboring nodes waste a 
significant  portion of bandwidth and incur higher delays to generate a TDMA-schedule. However, to 
establish a TDMA-schedule, if an 
algorithm incurs significant volume of message exchange and thereby consume more energy, and  this
may lessen the energy saving benefit  of using TDMA-based channel access. The 
problem of message overhead due to message exchange between neighboring nodes, becomes more severe 
for large and  dense networks. Therefore,  TDMA-scheduling algorithm with lesser overhead, not only 
save the channel bandwidth and reduce the time to generate a valid TDMA-schedule, such algorithms  
show better support for scalability too.

\item \textbf{Fairness:}
One of the crucial requirements of WSNs applications, is to maintain the fairness between the nodes  in terms of the opportunity to transmit their data.  For example, to get a consistent 
view of the sensed environment in WSNs, the sensor nodes should get equal opportunity to transmit their sensory data. 
Maintaining fairness is essential, especially for those applications in which the reading of each sensor node is equally important.
 However, many times in WSNs, few nodes in the networks 
also work as routers helping other's data to reach the destination. In order to ensure fairness, in this situation, 
such router nodes should get more number of time-slots than the nodes which are not the routers. The actual number of time slots 
required by a router depends upon number of nodes, from where it is receiving the packets to be forwarded.
\end{itemize}

\begin{figure*}[t]
\centering
\includegraphics[scale=.57]{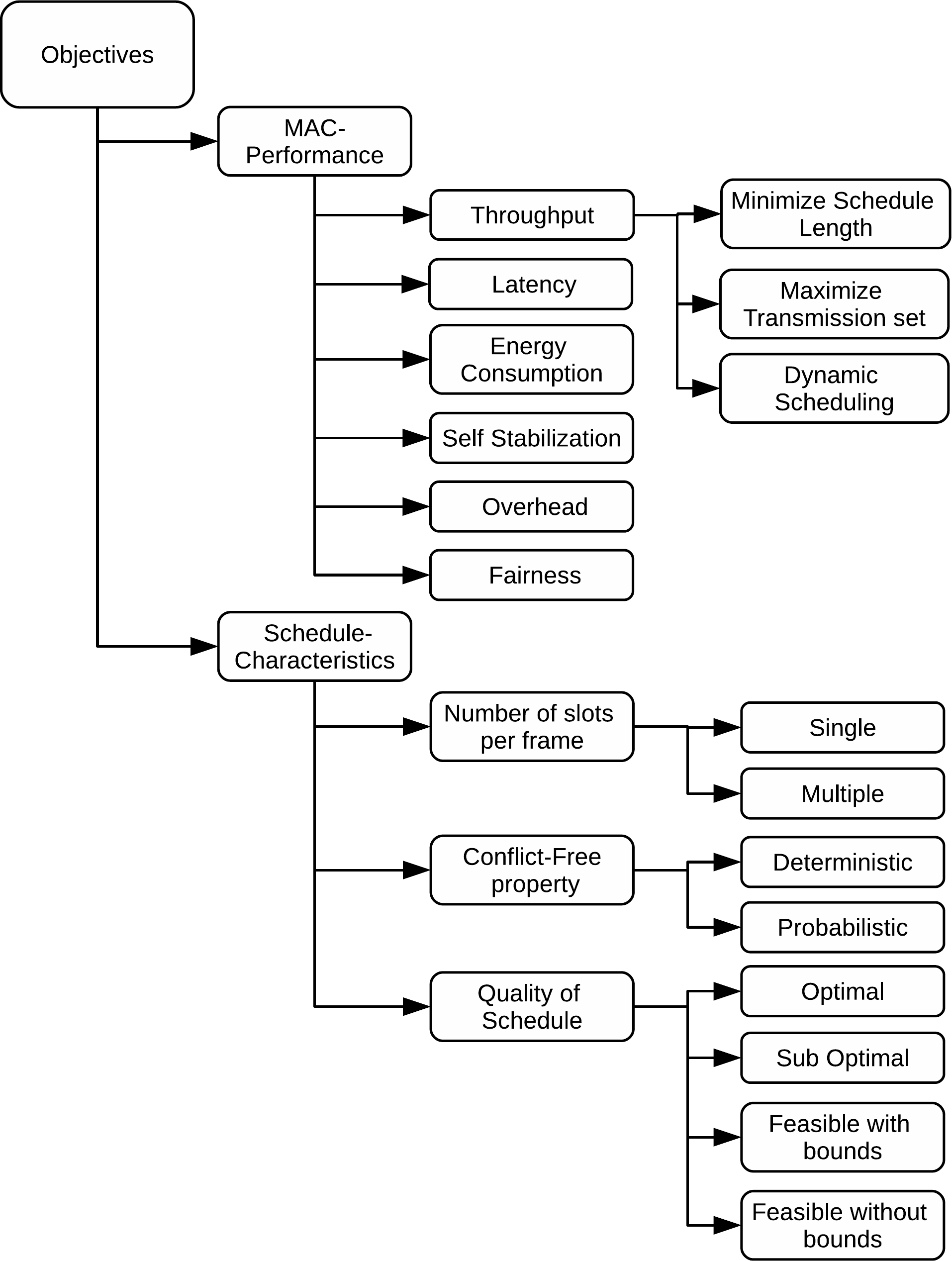}
  \caption{Classification of TDMA-scheduling algorithms based on scheduling objectives}
\label{fig:classification-obj}
\end{figure*}

\subsubsection{Schedule-characteristics objectives}
TDMA-schedules generated by two different algorithms can differ with respect to various 
characteristics,  even if both the algorithms are having the same MAC-performance 
objective. The are quiet a few reasons behind calling schedule-characteristics as  design objective of the TDMA-scheduling 
algorithms, which are as following. 

\noindent \texttwelveudash \hspace{.2cm} Such  characteristics of the schedule are usually decided 
before commencing the design of   algorithms. 

\noindent \texttwelveudash \hspace{.2cm} These schedule-characteristics greatly affect the design-methodology (discussed later) to 
be taken to perform scheduling. 

\noindent \texttwelveudash \hspace{.2cm}  These schedule-characteristics possesses a strong correlation with 
the MAC-performance objective of the algorithm.

\noindent  Following are some of the schedule-characteristics identified by us.

\begin{itemize} 
 \item \textbf{Number of slots per node:} Typically, the schedule generated by a TDMA-scheduling 
algorithm contains a  single slot per frame for each node in the network. This type of scheduling is 
useful for the case when all nodes in the network are having similar data transmission requirement, 
which may not be true all the time. For example, in WSNs, the sensor nodes which also work as 
routers to forward others data in the direction of base station, have more  data transmission 
requirement than that of sensor  nodes which only transmit their own data. Therefore, in order to 
ensure fairness, a couple of TDMA-scheduling algorithms allocate the number of slots to the 
nodes, in a frame, as per their data transmission requirement. In another situation, few nodes
may be given more than one time slot per frame to reduce the average waiting time of the packets to be transmitted,
in the MAC queue.

\item \textbf{Conflict free property:} As stated before, the task of a TDMA-scheduling algorithm 
is to generate a conflict-free schedule. However, there are couple of algorithms present in 
literature, in which, the purpose of scheduling is to avoid the collision to a certain extent, 
instead of completely suppressing it. This type of algorithms, usually combine the heuristic  scheduling and randomized transmission   to avoid possible packet collisions. This incurs lesser message overhead as compared to 
the message overhead required to generate a completely conflict-free schedule. However, the probabilistic nature of the schedule does not 
guarantee collision-free transmissions.

\item \textbf{Quality of schedule:} This objective can be seen as the level of  MAC- 
performance objective that has to be achieved by a TDMA-scheduling algorithm. In this regard, a TDMA-schedule can 
either be optimal, sub-optimal or feasible (with bounds or without bounds). Finding an 
optimal schedule is typically a hard problem with respect to many MAC-performance objectives. 
In general, the TDMA-sche-duling algorithms which  generate optimal or sub-optimal 
TDMA-schedule, have a single MAC-perfor-mance objective. On the other hand, the 
TDMA-scheduling algorithms with feasible quality-of-schedule objective, try to find a schedule which can 
improve the performance of MAC-protocol  to a certain extent, but not necessarily attain optimal or 
sub-optimal performance. Typically, the algorithms with feasible quality-of-schedule objective,
try to improve the MAC-performance with respect to multiple MAC objectives.
\end{itemize}

\subsection{Assumptions}
\begin{figure*}[t]
\centering
\includegraphics[scale=.45]{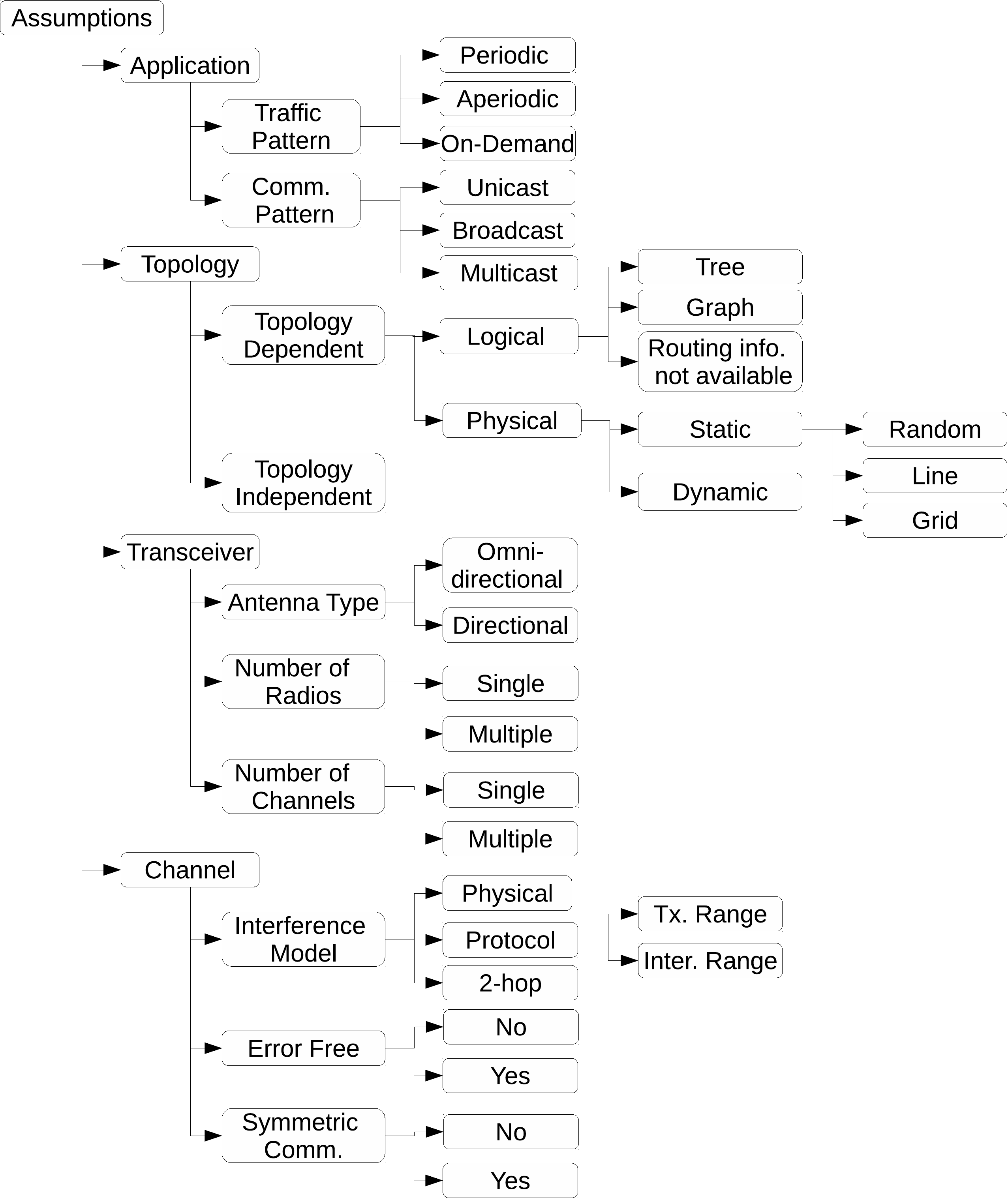}
  \caption{Classification of TDMA-scheduling algorithms based on assumptions made by the algorithm}
\label{fig:assumptions}
\end{figure*}

Every TDMA-scheduling algorithm, either explicitly or implicitly, define an underlying network  
model, on top which, the proposed algorithm is supposed to run. This is done by making certain 
assumptions about the parameters related to various aspects of the network (e.g. network topology, antenna type, 
node mobility etc.). Usually a TDMA-scheduling algorithm performs better in the network scenario, where most of the
assumptions made by it are satisfied. Moreover, there are certain assumptions which solely defines the objective of the 
scheduling and design methodology to be adapted. For instance, if nodes in the network are highly mobile then self-stabilization becomes an important objective to be achieved and the algorithms in which scheduling decision 
is taken at a centralized node, are not suitable. Similarly, few algorithms assume that the nodes are aware of 
their relative position, and therefore such algorithms can not be used at all, when such  assumption is not 
satisfied. Here, we categorize all the assumptions made by existing TDMA-scheduling 
algorithms, into following four major heads (Fig. \ref{fig:assumptions}). 
\begin{enumerate}
 \item Application 
 \item Network Topology and Routing
 \item Transceiver (Radio)
 \item  Channel (Transmission medium)
\end{enumerate}

\subsubsection{Applications}
The applications which are expected to be run on WSNs, can be characterized by making two 
types of assumptions: 

\begin{enumerate}
 \item \textbf{Traffic Pattern:}
The traffic pattern generated by various applications in WSNs networks can either be  periodic, 
aperiodic or on-demand. In case of periodic traffic, a node generates a fixed number of packets 
per frame. Usually, it is assumed that the size of packet is also fixed, and therefore, in addition 
to periodic traffic it also becomes constant bit rate (CBR) traffic. Most of the data collection 
applications in WSNs, fall in this category.  However,  it is also possible that the 
periodicity of packet generation is not same across all the nodes in the network. Some nodes may 
require more than one time slots per frame to transmit their packets, while  others  may not have 
sufficient number of packets to be transmitted in every frame. In this case, using the same TDMA-schedule for 
a long duration may result in the wastage of bandwidth. In aperiodic or Variable Bit Rate (VBR) 
traffic, the number of packets generated by a node per frame, is not fixed.  In this 
situation, dynamic scheduling techniques are more suitable as compared to static scheduling. 
Finally, in on-demand traffic  a sequence of packets is transmitted by the nodes, in response to 
some external or internal triggering. For example, in WSNs, a node starts transmitting packets in 
response to the  query received from the base station, or in case, it has detected some event of 
interest which needs to be reported to the base station.

\item \textbf{Communication Pattern:} 
From the link layer perspective,  three type of communication
patterns are possible: unicast, 
broadcast and multicast. In unicast communication, only a single node is the intended receiver of a 
transmitted packet, whereas in broadcast communication, all neighboring nodes of the sender are assumed to be the 
intended receivers.  Multicast communication is the generalized form of broadcast communication, where 
out of all  neighbor nodes, only the nodes belonging to a  predefined set (multicast group) are the intended receivers.
A TDMA-schedule, that has been generated for broadcast communication, may result poor bandwidth 
utilization, if the same schedule is used for unicast  communication too. Conversely, a 
TDMA-schedule, that has been  generated for unicast  communication,  may cause collisions of packets, if 
used for broadcast communication.

\end{enumerate}

\subsubsection{Network Topology}
Based on the network topology assumptions made by TDMA-scheduling algorithms, these algorithms can be 
classified into two categories: topology dependent and topology independent. The topology dependent 
algorithms assume that the nodes  have the  prior knowledge about the topology of the networks (such as 
size of the network and  membership of the nodes). Therefore, these
protocols are inappropriate for large networks or networks 
of varying size. On the other hand, the topology independent algorithms are 
transparent  of a specific topology and
therefore immune to node mobility. This makes the topology independent scheduling 
particularly attractive for mobile ad hoc networks. However, the bandwidth efficiency of a 
topology independent scheduling  is lower than that of a topology dependent scheduling due to its 
redundancy requirement, in order to work without topological information. Additionally, the efficient operation of 
topology independent schedule also requires an instant feedback channel which may not be available all the time. Therefore, 
the topology independent scheduling is not always applicable. 

The topology dependent scheduling can be 
further classified based on logical (routing) topology and physical topology assumptions made by the algorithms. 

\begin{enumerate}
 \item  \textbf{Logical Topology:} The logical (routing)  topology of a WSN
refers to the next-hop information  availability at the nodes that is established  by the underlying routing 
protocol. For example,  most of the applications in WSNs use tree-based routing topology for data collection.  
However, in general, TDMA-scheduling algorithms for  WSNs consider a generic 
graph as the underlying network topology. Additionally,  it is possible that a TDMA-scheduling 
algorithm does not make any assumption related to the availability of the routing information.

\item \textbf{Physical Topology:} This specifies the neighborhood relationship between the nodes of 
the network. This neighborhood relationship can either be static or dynamic based on the 
mobility model of the nodes.  In case of static topology, i.e., the node are not mobile, the physical topology 
of a network is mainly determined by the method of node deployment which usually varies with the application requirements. 
Random, line and grid are some of the most common methods of node deployment in WSNs.
\end{enumerate}

\subsubsection{Transceiver}
Typically,  sensor nodes contains a single channel  transceiver that at a time can be tube only to one frequency channel.  However,  many TDMA-scheduling algorithms have also considered the radios which can be tuned to more than one  channels simultaneously. In case of multiple channels, the TDMA-scheduling problem is often 
performed in conjunction with channel assignment problem, where the problems of 
assigning a communication channel and a time slot to a node are considered simultaneously.

\subsubsection{Channel}
The wireless channel is inherently erroneous. To support robustness, it becomes very 
important to consider the loss of protocol messages while designing a TDMA-scheduling algorithm. 
Although most of the existing algorithms consider the erroneous nature of wireless channel, there are few
algorithms which assume the channel as completely error free. Additionally, most of the 
TDMA-scheduling algorithms assume that the wireless channel is symmetric, i.e., the channel quality 
from node A to node B is same as that from node B to node A, which is not true all the time.

In wireless communications, the interference at is typically treated as  the sum of  signal levels present at the node due to 
all other unwanted transmissions.  The signal to noise ratio (SINR)  model and protocol model are two major ways to model the interference relationship between the nodes in a wireless network. The SINR model is also known as physical model \cite{Gupta}. 
According to the  protocol model, a message cannot  be received correctly if there is at least  one
sender other than the intended sender of a receiver is transmitting simultaneously  within its neighborhood. 
Typically, there are two approaches to estimate the range from the receiver within which no node should 
be allowed to transmit when the receiver is expecting a transmission: transmission range based approach 
and interference range base approach. Typically, the interference range is more than the transmission
range and gives better approximation for reality. However, estimating the transmission range is much easier than that of
estimating the interference range. One benefit  of protocol interference model is that, under this model,
it is easier to formulate the problem of TDMA-scheduling  as the  graph-coloring problem.  In \cite{Gronkvist}, Gronkvist et al. stated that the protocol interference model does not always give and accurate estimation of  interference present  between the nodes in reality as in wireless networks the actual level  interference at a node is a combined effect of  multiple nodes present in its proximity .

On the other hand, the physical model,  is better in the sense that it can measure the level of interference at a node accurately even in case when  multiple nodes are transmitting simultaneously in the proximity of the node. According to the physical model of the interference,  a message is said to be received successfully,  if the SINR at the receiver is more than  threshold. 
Moscibroda in  \cite{Moscibroda}  shows that protocols that are  designed considering the SINR based  interference model can even  perform the better than the theoretically achievable performance of graph-based scheduling protocols. Other than protocol and physical interference
modes, one more model, that is based on hop-counts, has been considered by  many existing algorithms. 
As per this hop-count interference model, two nodes can not take the same time slot if they are within the k-hop distant
from each other.

\subsection{Design Methodology}
\begin{figure*}[!h]
\centering
\includegraphics[scale=.55]{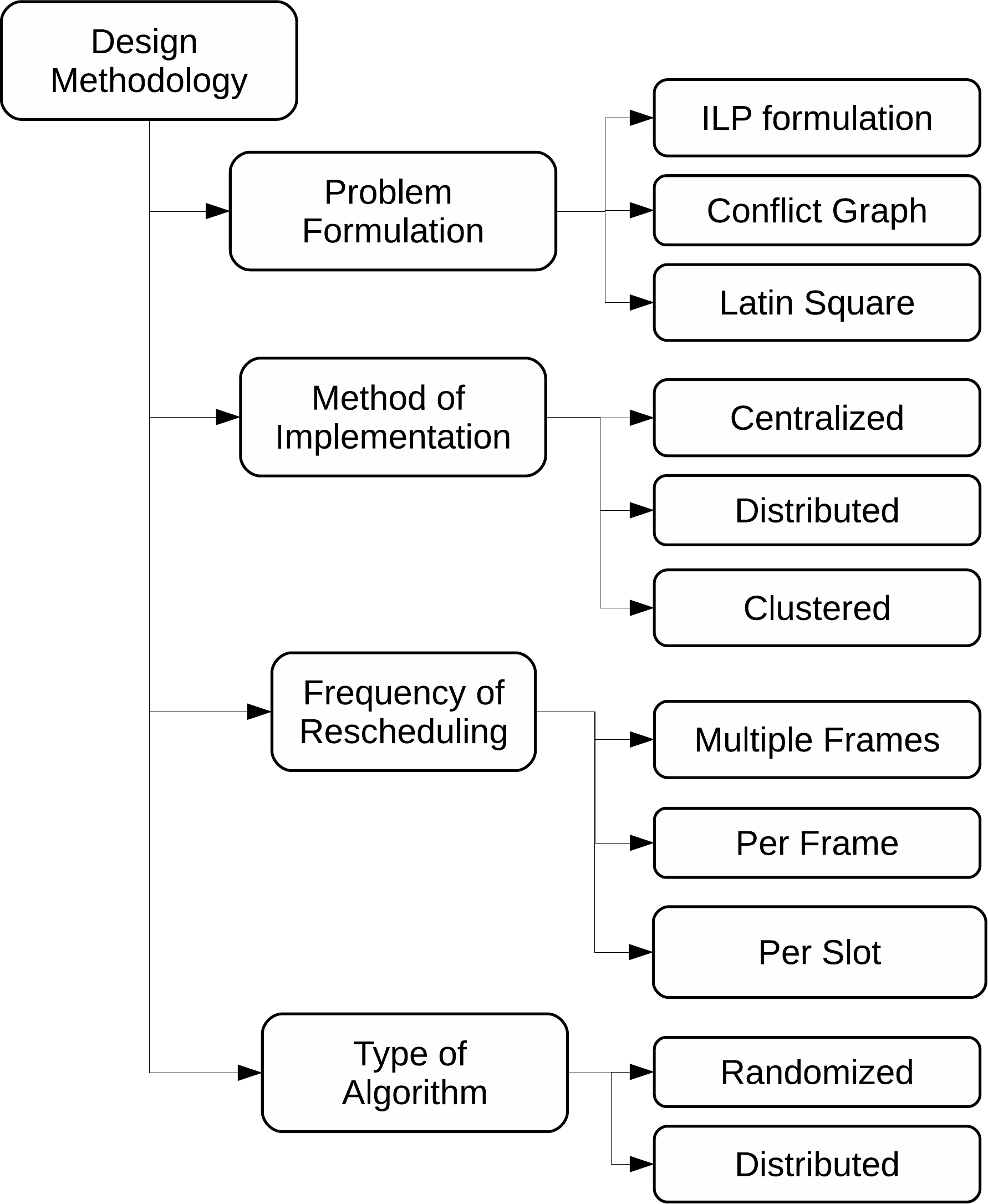}
  \caption{Classification of TDMA-scheduling algorithms based on design methodology}
\label{fig:design}

\end{figure*}

In a broader sense, the design methodology taken by an TDMA-scheduling algorithm is actually  the novel part of a 
TDMA-scheduling algorithm, where exactly, the contribution  of a proposed algorithm lies. Different 
TDMA-scheduling algorithms use different techniques to either achieve the same or different 
objectives. Two algorithms may differ, in terms of the scheduling techniques used by them at a detailed level. 
But, at the coarse level, they may be using similar concepts. For example, many TDMA-scheduling algorithms 
use classical graph coloring approach from graph theory to perform scheduling. However, the heuristics used by them
to decide the order in which the nodes in the graph would be colored, can be different.
Based on the higher level concept employed by various TDMA-scheduling algorithms, we further classify the design methodology 
taken by a TDMA-scheduling algorithms into following four sub-categories (Fig. \ref{fig:design}).

\begin{enumerate}
 \item Problem Formulation and Scheduling Technique 
 \item Method of implementation
 \item Frequency of rescheduling
 \item Deterministic vs. Randomized algorithm
\end{enumerate}

\subsubsection{Problem Formulation and Scheduling Technique}
\label{Problem Formulation}
Many scheduling algorithms use some mathematical construct to formally define
the actual TDMA-scheduling problem under consideration. Defining the problem in this manner 
works as the basic to theoretically analyze the correctness and performance of the algorithm, and also compare it with 
those of other algorithms which have considered similar problem.  Based on the problem formulation, a TDMA-scheduling 
algorithms can be categorized either as conflict-graph based, Integer Linear Programming (ILP) formulation based 
or latin square based.

\begin{itemize}
\item  \textbf{Conflict Graph:} A WSN can be considered as connectivity graph (
communication  graph), with  sensor nodes as the vertices of the graph.  There exist and edge between any two  vertices in this connectivity graph,  if  the sensor nodes corresponding to these vertices  can directly communicate with each other. 
Most of the TDMA-scheduling algorithm, formulate the  TDMA-scheduling problem using a 
conflict graph. 
The conflict graph can be considered as the line graph (conjugate graph) of a connectivity graph in such a way that every vertex of a conflict graph represents and edge of the the connectivity graph and two vertices the conflict graph are said to be adjacent if and only if their corresponding edges share a common vertex in connectivity graph.
Two nodes should not be assigned the same slot, if  the simultaneous transmission from these nodes causes the  interference at the receiver of 
either one of them.

After defining a conflict-graph  the TDMA-scheduling  problem is usually solved by using graph 
coloring approach. In its simplest form, the  graph coloring problem can be defined as the way of 
coloring the vertices/edges of a graph such that no two adjacent vertices/edges share the same 
color.  These algorithms typically use different heuristics (greedy approach) and various properties 
of underlying conflict-graph, to get the efficient TDMA-schedule.

\item \textbf{ILP formulation:} In this method, an Integer Linear Programming 
(ILP) formulation of the TDMA-problem is first provided considering resource and time constraints,  
and then it is solved for the  optimality. Some algorithms, also provide  a sub-optimal solution by solving 
the relaxed LP formulation of the original problem.  

 \item  \textbf{Latin Square:} Another technique employed  by a couple of TDMA-scheduling algorithms
is the use  of the Latin Squares (LS) characteristics \cite{Latin} to facilitate the assignment of time 
slots. An $n \times n$ latin square is a square matrix consists of numbers 1 to n arranged in a manner so that 
column or row contains the same number more than once.
 \end{itemize}

\subsubsection{Method of Implementation}
The method of implementation refers to the  place of scheduling control in a TDMA-scheduling  
algorithm where the  scheduling decisions can take place. According to this category, a 
TDMA-scheduling algorithm can either be centralized, distributed or clustered.

\begin{itemize}
 \item \textbf{Centralized:} In centralized TDMA-scheduling algorithms a single node (e.g. base 
station in WSNs) takes the 
responsibility to compute the  TDMA-sche-dule, and then, distribute it to all the nodes in the 
network. Such algorithms require the complete topology information to be available at 
base station, and therefore, these algorithms are suitable neither  for the large networks nor  for 
the networks in which the network topology changes frequently. However, the 
availability of complete topology information at a single point, allows these 
algorithms to generate optimal or sub-optimal schedule with respect to given design objective. 
Finding an optimal solution for most of the 
TDMA-scheduling  problems is NP-hard \cite{C-MSL-Ramanathan}, and therefore, these algorithms 
are not scalable in terms of processing time taken by the base station to generate an 
optimal schedule and  time required to collect the network topological information 
from all nodes present in the network. Furthermore, in centralized TDMA-scheduling, the 
schedule generated by the base station  is same for all nodes in the  network, and therefore,  the 
nodes need to be globally synchronized. 

In spite of aforementioned drawbacks, the centralized TDMA-scheduling algorithms are useful 
for small and static networks, and also help us to find out the theoretical bounds that can be 
achieved by any TDMA-algorithm for a particular objective. Furthermore, many times, the 
centralized algorithms often work as the seed towards the design of a distributed algorithm to 
solve the same TDMA-scheduling problem. 

  \item \textbf{Distributed:} While centralized algorithms rely on a single node to compute 
TDMA-schedule, in distributed 
algorithms, nodes compute their schedules by exchanging (implicitly or explicitly) the local 
information  with their neighboring nodes. The distributed algorithms support 
scalability for large networks, and also, they are adaptive to the dynamic changes in network 
topology. However, targeting for the optimal solution using distributed algorithms is not often 
feasible. This is  because, the  topology information of the complete network is not available at individual nodes. 
Therefore, the distributed algorithms try to generate either a feasible TDMA-schedule with no 
bounds 
or use some heuristic (greedy approach) to generate a sub-optimal schedule with guaranteed bounds. 
Additionally, in distributed scheduling, the nodes are required to be synchronized only with their 
neighboring nodes (local synchronization).  
  
   \item \textbf{Clustered:} The third type of TDMA-scheduling algorithms under this category are 
   cluster-based algorithms. In general, cluster-based TDMA-scheduling begin with the formation of clusters of  the network by 
selecting few nodes as the cluster heads, and associating the rest of the nodes to these
cluster heads. Thereafter, the cluster heads are responsible for generating the TDMA-
schedule among the nodes within their clusters. Cluster based algorithms prove to be
better scalable than the centralized algorithms. But, these algorithms often suffer from the problem of  inter
cluster interference due to intersection of  nodes  covered by  adjacent  cluster heads.  Additionally, the requirement of re-clustering due to frequent topology changes,
causes the cluster-based TDMA-scheduling algorithms not suitable for networks with dynamic topology.
 
\end{itemize}

\subsubsection{Frequency of Rescheduling}
Based on the frequency of rescheduling, a TDMA-scheduling algorithm can be classified either 
as static or dynamic. In static TDMA-scheduling, once a schedule is 
generated, the same schedule is used for sufficiently long time, without performing the 
rescheduling.  Usually, static TDMA-scheduling algorithms  generate optimal/sub-optimal schedule in terms of the 
schedule-characteristics objective, and the generation of  TDMA-schedule using such algorithm take very long time. 
On the contrary, in dynamic TDMA-scheduling  the  assignment of slots  is  performed  either on per slot basic or on  per frame basic.
Unlike static algorithms, the dynamic algorithms typically produce  a feasible TDMA-schedule in  very less time. These algorithm are good for the  situations where frequent  rescheduling has  to be performed 
 either  due to dynamic topology (logical or physical) or variable load conditions.

\subsubsection{Deterministic Vs. Randomized Algorithm}
A  deterministic TDMA-scheduling algorithm always  produces the same TDMA-schedule, for a specific combination of 
assumptions, objectives and underlying topology.
On the other hand, there are few TDMA-scheduling algorithms which use a degree of randomness (randomized algorithms) 
as part of their logic to generate a TDMA-schedule, and may generate different schedule for same input conditions,
every time it is executed.  Usually, the randomized algorithms achieve good ``average case'' 
performance, but sometimes these algorithms may produce an incorrect schedule or fail to produce a 
schedule within a bounded time period.

\section{Existing TDMA-scheduling Algorithms in WSNs}
In this section, we discuss existing work on TDMA-scheduling algorithms in WSNs particularly designed for the nodes 
with single radio, single channel, and with omini-directional antenna transceiver characteristics. The list of scheduling 
algorithms discussed here, is a subset of work on TDMA-scheduling available in the literature. However, the algorithms are 
chosen in such a way so that most of the characteristics presented in the classification framework proposed 
in section \ref{section:classification}, are covered. In the following, we first give the description of 
chosen algorithms followed by a summary of their characteristics in terms of scheduling objective, assumptions and 
design methodology.

The order of discussion for the selected TDMA-scheduling algorithm is as follows.  At the very first level, 
we  group all the algorithms, based on the method of implementation(centralized, distributed and clustered), 
and discuss the algorithms belonging to a 
particular method of implementation in a single subsection. At the second level, i.e., inside the discussion of 
the algorithms which belong to a particular method of implementation, we 
group the algorithms based on the scheduling objective, and discuss them together.

\subsection{Centralized TDMA-scheduling Algorithms}
Here, we discuss the centralized TDMA-scheduling algorithms available in the literature, and with  scheduling objectives such 
as maximizing throughput, minimizing latency, minimizing energy and maximizing fairness. Usually, minimizing overhead, 
and self organization are not the scheduling objectives of centralized TDMA-scheduling algorithms.

\subsubsection{Algorithms on Maximizing Throughput}
One of the early work in this category appeared in \cite{C-MSL-Ramanathan} in which  
the TDMA-scheduling problem is considered analogous to the vertex coloring problem in graphs. 
Consider the set of colours as integers ranging from 0 to  $\Delta$,  where $\Delta$ is the size of maximum distance-2 neighborhood in the graph.  Let the color assigned to two node $u$ and $v$ are $f_u $ and $f_v$ respectively. Then  $|f_v - f_w| > 1$, if $u$ and $v$ are neighboring nodes.

In this work \cite{C-MSL-Ramanathan}, the authors proposed three three different centralized algorithms for TDMA-scheduling. All three algorithms are based on breath-first approach and   differ with respect to the order the nodes are scheduled or colored. 
In the first algorithm, the next node to be assigned a color is  picket randomly from the set of all uncolored nodes. The second algorithm, colors the nodes as per the increasing order of the degree on nodes in the conflict graph representation of the network. The third is very much similar to the second with the difference that in this approach the nodes are removed from the graph after assigning them a  color.  The distributed implementation of the approaches two and three are not possible as the  
ordering requires global information. In all the approaches, the worst case coloring could be as high as  $\Delta$
(Vizing's Theorem \cite{Vizing}).

Florens et al. in  their works  \cite{C-MSL-Florens1}, \cite{C-MSL-Florens2}, \cite{C-MSL-Florens3}, propose a couple of  centralized TDMA-scheduling  algorithms especially for packet transmissions from base station to  the sensor nodes.  The objective of these is to  algorithms is minimize the schedule-length for line topology, assuming  protocol interference model. 
The basic approach behind the scheduling is to  transmit packets to the nodes for the number of hops from the base station is more than the others

By reducing the minimum schedule length problem to the graph coloring problem, Ergen et al.  \cite{C-MSL-Ergen} showed that the schedule length problem is also NP-complete.  Intuitively,  the reason behind scheduling problem of becoming an NP-complete problem is due to the fact that the assignment of a time slot among many candidates effects the eligibility of many other nodes to take some other slot simultaneously.  Additionally, many eligible nodes may not have data to transmit in a slot as a consequence of the nodes selected in previous slot.

In order to achieve uniform flow of data in a network,  the authors of \cite{C-MSL-Ergen} propose a disjoint path based approach which works by constructing multiple disjoint equally spaced paths from nodes to the base station. Two different path strips  can maintain two different data simultaneously  without interfering each others transmission.  This allows uniform load distribution among the nodes in the network and hence  prevents prevents a small set of nodes from being overloaded.It is shown that, in contrast to the traditional shortest path routing, the schedule generated by path strip based approach is smaller than the schedule length generated by the shortest path routing.

Choi et al. \cite{C-MSL-Choi} combined the optimal scheduling problem and routing problem together for the scenario where every node has only one packet to send. They proved that the  problem  of scheduling and routing together is NP-complete, and also proposed algorithms which generate  schedule with less than 3N-3 time slots for line and tree topologies. For general graphs,  they proposed a  heuristic based solution by first creating a  minimum spanning tree and then removing  its edges one by one  edges so that simultaneous transmission in two different branches do not create  interference for each other.

Annamalai et al. \cite{C-MSL-Annamalai} studied the use of  orthogonal  codes such as FHSS and DSSS to mitigate the 
interference between the nodes sharing same time slots. They proposed a top-down tree construction scheme based on a
greedy approach in which the children of a nodes is selected as nearest neighbor by traversing the tree in BFS 
(Breath First Search) order.  To mitigate the interference,  the child nodes of a parent, 
are assigned different codes.  In case,  new code is not available,  then assign the code that least used by the 
neighbor nodes.  After code assignment, the procedure for time slot assignment is performed in a manner such that a
parent gets the slot  after its children only.

\subsubsection{TDMA-Scheduling Algorithms on Minimizing Latency}
Cui et al \cite{C-ML-Cui} looked the problem of  TDMA scheduling with the objective to minimize the end-to-end latency 
of involved for convergecast communication, instead of minimizing the schedule length. They also analyzed the trade-off
between energy consume by the nodes and  the latency of data transmission for the scenario when the number of packets transmit by each node, in a time frame,  is same across all the nodes in the network.
The authors prove that   in order to  achieve the minimum possible latency,  it is sufficient to schedule the outgoing links of every node after scheduling  their  incoming 
links The proposed algorithm first divides the links into different levels based on the number of hops they are far from the 
sink,  and then create a  schedule  beginning from farthest node to the nearest node.

Instead of defining a TDMA schedule by assigning time slots to the senders, the work proposed in \cite{C-ML-Cui}
generates a receiver based schedule, where the slots are assigned to the receivers. The receiver based scheduling
is particularly useful for the devices which perform duty cycling (sleep scheduling) and wake-up only when they intend to 
either receive or transmit a packet. The authors first proved that deciding the frequency of beacon transmission in 
a beacon-enabled

Revah et al. in \cite{C-ML-Revah} argue that the TDMA-scheduling techniques with the aim of minimizing  
 the  convergecast  completion time  do not consider the waiting time of messages. It is  not reasonable to hold a  message  if  the time slots are available to transmit the message  immediately. The authors proposed a number of  algorithms for different network topologies such as  linear, two-branch, and star. One of the major assumptions made by these algorithms is the availability of  different directional radios for  upstream and downstream control channels.

Lu et al. in \cite{C-ML-Gang}  studied the problem of routing and scheduling jointly with the objective of minimizing transmission latency. Given a graph, a number of flows and slots,  find  paths  and a slot assignment such that 
it  minimizes average latency with maximum number of flows. A graph coloring approach is used to solve the problem. 

\subsubsection{TDMA-Scheduling Algorithms on Minimizing Energy}
In \cite{c-ME-Kalpakis}, Kalpakis et al. consider the problem of TDMA-scheduling with the objective to maximize the lifetime of  WSNs. In particular, they considered the network lifetime as the time until the first sensor node in the network runs out of energy. They propose an iterative algorithm to find the schedule which maximizes the lifetime of the network according to above definition. The problem formulation is done as the network flow problem and solved it  using ILP.

In \cite{c-ME-Mao},  the authors looked the problem of TDMA-scheduling  in WSNs with two different perspectives 1) minimizing the energy consumed by the sensor nodes due to excessive switching between active and sleep modes 2)  minimizing the total time to collect the data at the base station.  To solve the  optimization problem, the paper combines two different stochastic optimization techniques: the genetic algorithm and the swarm optimization. The swarm optimization algorithm ensures that there is no empty slot in the resultant schedule, whereas the genetic algorithm solves the optimization problem in lesser time. 
The authors showed a considerable improvement due to proposed mixing of two algorithm over particle swarm optimization algorithm alone, in terms of length of the generated schedule and the energy consumption due to excessive radio  stitching.

\subsubsection{TDMA-Scheduling Algorithms on Maximizing Fairness}

In \cite{C-MF-Sridharan} authors gave a linear programming formulation for  fair TDMA-scheduling problem in WSNs and based on the given formalism also propose a scheduling mechanism on a data gathering tree. The algorithm proposed in  \cite{C-MF-Sridharan} works in multiple rounds where after every round the nodes increase their data rate by a small value $\epsilon$, till the total bandwidth usage reaches  total available bandwidth.

Chatterjea et al. in  \cite{C-MF-Chatterjea} introduced AI-MAC as the extension to a schedule based MAC protocol LMAC \cite{LMAC}, to ensure the fairness among participating nodes. The AI-MAC differs from LMAC in the sense that in LMAC protocol each node is allocated only one slot in a frame whereas in AI-MAC a node can have multiple slots in a frame.The number of slots given to a node in particular frame depends upon the traffic load at the node. This ensures fairness among the nodes.
Table \ref{table:summary-centralized} provides the summary of surveyed centralized TDMA-scheduling algorithms 
as per the classification presented in section \ref{section:classification}.

\subsection{Distributed TDMA-Scheduling Algorithms}

\subsubsection{Algorithms on Maximizing Bandwidth by Minimizing Schedule Length}

Based on the centralized scheduling scheme RAND \cite{C-MSL-Ramanathan}, Rhee et. al. \cite{D-MSL-Rhee} proposed a distributed randomized time slot scheduling algorithm (DRAND). The authors of DRAND have also  used this algorithm  in  Z-MAC \cite{D-MSL-Rhee-ZMAC} protocol for sensor networks, to improve the performance of MAC protocol by leveraging  the  strength of TDMA-based channel access and contention-based channel access mechanisms simultaneously.

The execution of DRAND algorithm happens in multiple round where  a node passes through with four different sates namely IDLE, REQUEST, GRANT, and RELEASE   during each round. The duration of each round is not fixed and depends upon the estimate of the communication  delays between the nodes in the network.  In IDLE state a with probability 1/2 runs a lottery which further has some probability of wining. Then the node starts negotiating a time slot with its neighbors only when it wins the lottery and enters the REQUEST state. Similarly, the node which grant a particular time-slot to a requesting node enter into the GRANT state.  These state transitions finally reach to a  conflict free TDMA-schedule. Due to large message delays, the  runtime complexity of DRAND algorithm  increase rapidly with respect to increase in the density of the network $\delta$, where $\delta$  is defined as the average number of  nodes in a two-hop neighborhood of the network.

Ashutosh et.  al.  in  \cite{RDTDMA},  proposes  a distributed and randomized  algorithm (RD-TDMA) for TDMA-scheduling  based
on graph colouring approach. A major advantage of RD-TDMA
algorithm over other distributed TDMA-scheduling algorithms, such as DRAND,
is the multi fold reduction in scheduling-time by allowing all the nodes to
concurrently select their slots using probabilistic approach. This
is because, the static TDMA-scheduling algorithms typically
use heuristic based approach for graph
colouring that is essentially sequential in nature.

The execution of RD-TDMA  algorithm happens in multiple round. During the process of scheduling,  a  node passes through with four different sates namely Contention, Verification, Scheduled and Termination. Every node starts the algorithm by entering in the contention state. At this state and at time slot $s$, a node $i$ tries to take the slot $s$ with probability $\boldmath p_{i,s}$ by broadcasting request message to its neighbor nodes and enters into the verification state to check if any other node is also trying to get the same time-slot. If  yes, the node $i$ comes back to the contention state and starts all over again. Otherwise, if no node other than $i$  is trying for the same slot $s$ (received grant message from all other nodes), the node $i$ would take the slot $s$ and enter into the scheduled state. At this state, the node $i$ is not sure that all other nodes are also being informed that the slot $s$ has been taken by node $i$. Once all the neighbors of node $i$ know that node $i$ is in scheduled state, the node $i$ will enter into the termination state. At the end of the algorithm execution, all the nodes will be in termination state and will have some slot in a frame.  Later  the authors of RD-TDMA proposed another TDMA-Scheduling algorithm called DTSS \cite{DTSS} to extend the scope of RD-TDMA algorithm from only  broadcast scheduling to unicast, 
multicast and 
broadcast modes of transmission.
However, the DTSS algorithm requires global time synchronization among the nodes even before start executing  the algorithm.

The protocol in  \cite{D-MSL-Kevin} proposes a slot assignment algorithm based on the heuristic that is used to choose  the order to assign the slots to sensor nodes. Comparing to DRAND algorithm, the time complexity of this  algorithm is  larger, however the length of  schedule generated by the CCH  is smaller than that of DRAND.

Ranjeet et. al. \cite{D-MSL-Ranjeet} proposed a  TDMA slot assignment algorithm for WSNs 
which assumes the presence of a special node 
called mobile agent in the network. 
Mobile agent is responsible to allocate the time slots to the  nodes  which is does  by reaching near to the individual nodes present in the network.  Similar to the  greedy graph coloring  algorithm, the  mobile agent  upon receiving a query from a node, assigns the smallest available time slot to the node so that the node can use that time slot to transmit its data without any conflict. 
The time taken by the algorithm is constraint by the speed of mobile node, and therefore this solution is not suitable for networks
covering large geographical area. Additionally, the complete rescheduling needs to be performed every time the 
topology of the network changes.

Ashutosh et. al.  in \cite{DSLR}  proposes the idea of  iteratively reducing  the schedule length of an  existing  TDMA-schedule.
The algorithm for  reduction of schedule-length runs in rounds so that at the end of each round a valid TDMA-schedule is generated. 
In this way, the algorithm always produces a valid  TDMA-schedule even if it is stopped  after the  arbitrary number
of round of execution. 
 In this way,   a TDMA-schedule of desired length can be generated by executing the algorithm for a limited number of rounds, thus  providing the ability to trade-off between schedule-length and the time to schedule.    In order to compress the schedule-length, all 
the nodes shift to a  free slot that comes before the slot currently occupied by the node,
without violating the conflict-free property of existing
schedule . The algorithm ensures that during the process of shifting from higher Id slots to lower Id slots,  the neighboring nodes do not shift to the same free-slot simultaneously.

\begin{landscape}
\def\baselinestretch{1}
\footnotesize
\tablefirsthead{
  \hline
   \textbf{Ref.} & \multicolumn{4}{|c|}{\textbf{Objective}} & \multicolumn{5}{|c|}{\textbf{Assumptions}} & 
   \multicolumn{3}{|c|}{\textbf{Design Methodology}} \\
  \hline
        & \textbf{MAC Performance}  &  \multicolumn{3}{|c|}{\textbf{Schedule- Characteristics}} &  \multicolumn{2}{|c|}{\textbf{Application}} &  
        \multicolumn{2}{|c|}{\textbf{Topology}} &   
\textbf{Inter. Model} &  \textbf{Prob. Formulation} & \textbf{Freq. of Resche.}  & \textbf{Algo. Type}   \\
   \hline
     &  & \textbf{Slots/ Node} & \textbf{Conflict Free} & \textbf{Optim- ality} &  \textbf{Traffic Pattern} & 
     \textbf{Comm. Pattern} & \textbf{Logical Topo.} & \textbf{Physical} &  
   & & & \\
   \hline }

\tablehead{
 \hline
   \textbf{Ref.} & \multicolumn{4}{|c|}{\textbf{Objective}} & \multicolumn{5}{|c|}{\textbf{Assumptions}} & 
   \multicolumn{3}{|c|}{\textbf{Design Methodology}} \\
  \hline
        & \textbf{MAC Performance}  &  \multicolumn{3}{|c|}{\textbf{Schedule- Characteristics}} &  \multicolumn{2}{|c|}{\textbf{Application}} &  
        \multicolumn{2}{|c|}{\textbf{Topology}} &   
\textbf{Inter. Model} &  \textbf{Problem Formulation} & \textbf{Frequency of Resche.}  & \textbf{Type of Algo.}   \\
   \hline
     &  & \textbf{Slots/ Node} & \textbf{Conflict Free} & \textbf{Quality of Schedule} &  \textbf{Traffic Pattern} & 
     \textbf{Comm. Pattern} & \textbf{Logical Topo.} & \textbf{Physical} &  
   & & & \\
   \hline
      }
   
\tabletail{%

\hline

\multicolumn{13}{|r|}{\small\sl continued on next page}\\

\hline}

\tablelasttail{\hline}

 \topcaption{Summary of Centralized TDMA-scheduling algorithms in WSNs}
\begin{supertabular}{|p{1.3cm}|p{1.0cm}|p{.8cm}|p{1.1cm}|p{1.0cm}|p{1.0cm}|p{1.1cm}|p{1.2cm}|p{1.1cm}|p{2.6cm}|p{2.0cm}|p{2.4cm}|p{.7cm}|}
\label{table:summary-centralized}
\cite{C-MSL-Ramanathan} &  MSL  & single & complete & sub optimal & Periodic & Broadcast & Graph 
& Random &  2-hop, protocol(Tx range) & Conflict Graph  & Multiple Frames & Det.  \\
\cite{C-MSL-Florens1,C-MSL-Florens2,C-MSL-Florens3} &  MSL & single & complete & optimal & 
Periodic & Unicast & line \& multiline & Random &   protocol (Tx range) &  Conflict Graph  & 
Multiple Frames & Det.   \\
\cite{C-MSL-Ergen,C-MSL-Annamalai,C-MSL-Malhotra} &  MSL & single & complete & optimal 
& Periodic & Unicast & tree & Random & protocol (Tx range)&  Conflict Graph  & 
Multiple Frames & Det.   \\
\cite{C-MSL-Choi} &  MSL & single & complete & optimal 
& Periodic & Unicast & tree & Random &  2-hop (Tx range)&  Conflict Graph  & 
Multiple Frames & Det.   \\
\cite{c-ML-gandham} &   MSL & single & complete & optimal 
& Periodic & Unicast & line & grid & protocol (Tx range)&  ILP & 
Multiple Frames & Det.  \\
\cite{C-ML-Cui} &  ML & single & complete & optimal 
& Periodic & Unicast & tree & Random &  2-hop (Tx range)&  Conflict Graph  & 
Multiple Frames & Det.   \\
\cite{C-ML-Pan} & ML & single & complete & optimal 
& Periodic & Unicast & tree & Random & 2-hop (Tx range)&  Conflict Graph  & 
Multiple Frames & Det.   \\
\cite{C-ML-Revah} &  ML  & single & complete & optimal 
& Periodic & Unicast & Line, Star  & Random &  protocol (Tx range)&  Conflict Graph  & 
Multiple Frames & Det.   \\
\cite{C-ML-Gang}   &  ML  & single & complete & optimal 
& Periodic & Unicast & tree & Random &  protocol (Interference range) &  Conflict Graph  & 
Multiple Frames & Det.   \\
\cite{C-MF-Sridharan} &  MF & single & complete & optimal 
& Periodic & Unicast & Graph  & Random & protocol (Interference range) &  Conflict Graph  & 
Multiple Frames & Det.   \\
\cite{C-MF-Chatterjea} &  MF & single & complete & optimal 
& Periodic & Unicast & Graph  & Random &  protocol (Tx range) &  Conflict Graph  & 
Multiple Frames & Det.   \\
\cite{c-ME-Kalpakis,c-ME-Mao}  &   ME &  single & complete & optimal &  Periodic & Unicast & Graph  
& Random & protocol (Tx range) &  ILP  & Multiple Frames & Det.   \\

\end{supertabular}
\vspace{5pt}

\noindent \textbf{MSL: Minimizing Schedule Length, ML: Minimizing Latency, MF:Maximizing Fairness, ME: Minimizing Energy}
 \end{landscape}

\vspace{10pt}
\normalsize

\subsubsection{Algorithms on Maximizing Bandwidth by Maximizing Transmission Set}

Rajiv et al. in \cite{D-MSL-Rajiv},  addressed the problem of TDMA-scheduling for  broadcast communication in radio networks, emphasizing the fact that the wireless 
is inherently broadcast and many applications such as distributing updated database, routing tables etc make use use of this property.
In order to efficiently utilize the bandwidth, they defined broadcasting-set as the set of nodes that can broadcast in the same
slot without conflicts. A maximal broadcasting-set is then defined, as the broadcasting-set such that if any node is added to 
this set,  it becomes no longer a broadcasting set. They proved that finding out a maximum broadcasting-set is NP-complete. 
The scheduling algorithms presented in \cite{D-MSL-Rajiv} runs in two phases. 
During the first phase a broadcast frame is produced where each node is scheduled in exactly one slot per frame.
During the second phase, the algorithm produces a maximal broadcasting-set in each slot of the frame and it is done as follows.

A source node generates and broadcasts a token message.  The path taken by the token is the DFS of the graph. On confirming that, 
all the neighbors have received the token,  the source node selects its transmission  slot. Along with the token, the source node also  broadcasts  its current schedule table. On receiving the schedule table from the source node the neighboring nodes update their own schedule table, and also 
decide their new schedule based on this information.

A similar solution for broadcast scheduling is given by \cite{D-MSL-Ephremides}, where the algorithm 
starts with a skeleton schedule of N nodes for which the $i^{th}$ slot is reserved for the $i^{th}$ node.
By broadcasting their  schedule  neighbors  each node comes to know the information  about the unreserved slots which it can 
pick without creating any conflict with other nodes. However, to ensure that only a unique node should pick a slot 
in a two-hop vicinity, the algorithm uses node Id to decide the priority among the candidate nodes, instead of passing 
a token message  to each node one by one, similar to the work presented in \cite{D-MSL-Rajiv}.

\subsubsection{Algorithms on Maximizing Bandwidth by Dynamic Scheduling}
Sidi and Cidon \cite{D-dynamic-Cidon}  proposed distributed and dynamic 
link-scheduling algorithms for multi-hop packet radio networks. The dynamic nature of the algorithm achieves higher 
slot utilization  in case of topology and traffic changes.
The algorithm divided the shared channel into control segment and the data segment. The control or signal information among the nodes can only be transmitted during the control segment. The control segment is further divided into two segments: request segment and the confirmation  segment. 
Both the request and confirmation segments are further divided into N time-slots, where N is the number of nodes in the network.
When a node $i$ wants to reserve a time slot in the information segment, it transmits a request signal at the corresponding 
control segment. If node $i$ does not hear a deletion signal from any of its neighbor, it transmits a confirmation signal
and transmits a packet during the information slot. 
Treating the assignment problem independently has the fairness  issue as few nodes may get frequent permission to transmit their data than others.  Additionally, overhead (in terms of mini-slots) due to control signals 
per slot is the order of number of nodes in the network.

A distributed heuristic based  TDMA slot assignment algorithm  FPRP,  based  on  reservation 
cycle among the nodes in a two-hop neighborhood,  is proposed in \cite{D-dynamic-zhu}, where a cycle consist of five different phases.  The FPRP divides the time into  series of time frames with each frame consisting of  reservation and data transmission portions.
A node wanting to send the data has to first reserve a slot in the next time frame using the reservation portion of the current time frame. The nodes can use contention-based channel access mechanism in the reservation portion.  
The FPRP runs a five-phase cycle multiple times to  decide the winner of a time-slot.
In addition to distributed algorithm, the FPRP is a parallel protocol in the sense that multiple reservations may be made 
in parallel across the network. 
The algorithm makes two major assumptions that while  in receiving mode,  a node is able to tell whether one packet,  multiple packets  or no packets at all, were transmitted. To a large extend, this assumption is not valid in case of WSNs.

The work in \cite{D-dynamic-Chipara}  presents  a TDMA-scheduling  algorithm specifically  for query based  data aggregation in WSNs.  The algorithm assumes that the sensor nodes are supposed to report data periodically to the based station.
 The base station constructs a routing tree to disseminate a query to collect the data from sensor nodes. In the reverse path, each internal node aggregates the data received from its children and forwards it to its parent over the rooted tree.
The basic idea of the proposed scheduling algorithm is to exploit the precedence constraints imposed by the 
query path and data aggregation.

Salonidis et al. \cite{D-dynamic-Salonidis} propose 
distributed dynamic scheduling algorithm  for tree based logical (routing) topology in ad-hoc networks.
This proposed algorithm  starts from a initial T	DMA-schedule and iteratively reduces the schedule-length as per the demand set by higher layers. Additionally, unlike 
\cite{D-dynamic-Chipara} the algorithm proposed in \cite{D-dynamic-Salonidis} does not assume global clock synchronization.
In \cite{D-dynamic-Salonidis}, every link of underlying routing tree link is characterized by a slot demand.  
A  node initiates a rescheduling procedure asynchronously  when the application layer changes the demand 
 The  nodes reach to a valid TDMA schedule starting from the 
current  TDMA schedule using only local information.

\subsubsection{Algorithms on Minimizing Overhead}

The work in \cite{D-MSL-Wang} proposes a  deterministic distributed TDMA-scheduling algorithm (DD-TDMA) for
WSNs, while keeping the primary objective of generating a bandwidth efficient schedule, 
In  DD-TDMA a node decides its
own slot according to the slot occupancy status of its two-hop neighboring nodes. DD-TDMA scheduling is based on the assumption that the receivers can detect the collision i.e., in case of collision,  even if a receiver cannot correctly receive the packet,  it can  at least detect that some transmission has happened. 
This property removes the need to wait for an acknowledgment from neighboring nodes to  mitigate the collision.
After taking a slot, the node updates its one-hop neighbors with this information by broadcasting a message, which is then forwarded by the one-hop neighbors of the node to make the assignment information reach at the two-hop neighbors of the node.  The above  process is  repeated in 
every frame until finally all nodes are scheduled.

A MAC protocol called SEEDEX for ad-hoc networks proposed in \cite{D-Overhead-Rozovsky} tries to avoid 
collision without making explicit reservation for each and every packet. 
At the beginning of each slot, if a node has a packet ready for transmission, it chooses a slot with probability 
$p$ in which it can possibly transmit a packet (state PT), otherwise it will stay silent with probability $1 - p$ in
that slot (state 
Listen). Suppose a node A has a packet to transmit to a neighboring node B, then, first, the node A waits 
for a slot at which simultaneously node A is in PT state and node B is in listen state. At such a slot, node A may discover 
that there are $n$ other nodes neighbors of B which are also in PT state. Then node A transmits with  probability $1/n$ and 
refrain from transmitting packet in that slot with probability $ 1 - 1/n$. This technique is also called topology independent 
scheduling. 

The key concept of the proposed protocol lies in the fact that to know the status of their two-hop neighbors i.e.,  
whether they are in PT or listen state, the nodes simply exchange the seeds of their random number generators with their 
two-hop neighbors instead of  explicitly transmitting their status information. 
This approach considerably reduces the overhead due to message exchange. 
However, this technique mitigates the possibility 
of collision to a large extent, but does not guarantee a  collision free transmission.

A similar approach of using random seeds of neighboring nodes to determine slots is used in \cite{D-Overhead-Bao}.
The  algorithm proposed in  \cite{D-Overhead-Bao}  uses a hash function to determine 
priority among contending neighbors. When a  node $i$ wants to transmit in a slot $t$, it computers a priority $p_k^t$ for 
each member $k$ belonging to its two-hop neighborhood including itself, as: $p_k^t = Rand(k \oplus t) \oplus k$, where function 
Rand(x) 
is a pseudo random number generator that produces a uniformly distributed random number using the random see $x$. 
If $p_i^t > p_j^t, \forall k$ belonging to its two-hop neighborhood, then node $i$ can access the channel during slot $t$.
While the Rand function can generate the same number on different inputs, each priority 
number would be unique,  since $p_k^t$  is appended with k to the corresponding $Rand(k \oplus t)$.
A node being part of multiple overlapping neighborhoods may not take a slot even it is having the highest priority in one neighborhood but not in  others. In this way, the length of the schedule generated by the algorithm can go up to the number of nodes in the network leading to poor spatial reuse of the bandwidth. Also,  the computational complexity of the this type of scheduling is very high as  each node has to calculate the priority 
of all of its two-hop neighbors for every slot. The proposed scheduling assumes two-hop interference model where only one node is allowed to use a slot for transmission in a neighborhood. Although, two-hop interference eliminates the problem of hidden terminal but gives poor bandwidth utilization due to exposed node problem. 
Rajendran et al. \cite{D-Overhead-Rajendran}  proposed a distributed TDMA-scheduling algorithm as the successor of the algorithm proposed in \cite{D-Overhead-Bao}. in this extended protocol,  after calculating its own priority a node will 
announce the slots that it will use, a list of all  receivers for these slots and a list of slots  for which it has the highest priority
but it will not use. If a node choses not to use a slot for which it has highest priority the other nodes can use the same slot by again finding a winner ( a  node with highest priority) among themselves.

Lin et. al. \cite{D-Overhead-Chih-Kuang} 
propose  a distributed algorithm  for WSNs. The  algorithm generates  TDMA-schedule with high network utility 
analogous to DRAND. In addition to that,  it reduces the overhead due to protocol messages by
 exploiting the sensor location information.

\subsubsection{Algorithms on Self Stabilization}
In \cite{D-Stabilization-Ammar}, the authors have addressed the problem 
of rescheduling (self stabilization) in presence of mobile nodes.  In particular,
they proposed a procedure for TDMA-schedule restructuring to maximize the bandwidth utilization by utilizing the unused slots (secondary slot assignment) . The restructuring is performed in such a way so that it  involves the slot reallocation to the minimum number of  nodes. 
In  order to perform rescheduling, the nodes exchange control messages with their  neighboring 
nodes via a separate control channel in the form of an extra slot in the TDMA frame. 
The algorithm uses a concept of primary and secondary slots assignment for the nodes to transmit their data. 
The primary slots are given to the nodes which have recently moved to a new neighborhood. 

Ali et al. \cite{D-Stabilization-Ali} proposed  adaptive and distributed 
TDMA algorithms for multi-hop wireless  networks. One of the unique feature of the algorithm presented in \cite{D-Stabilization-Ali} is the method to 
detect the presence of a new node in dynamic topology when nodes are mobile. 	
In the absence of collisions, a node can detect the presence of a new node in its proximity 
just by receiving the normal packets transmitted by the new node. 
But, the transmissions of new node  cannot be received if there is an interference  at the  receiving node.  
To detect the new neighboring node in this situation, they utilized the initial portion of a time slot where a node transmits its own address before commencing the transmission of actual data. This is done by introducing an additional  field called flag field  to the packet header. A node may also decide (with probability 1/2)  not to transmit the header and keep listening during the initial portion of the slot. This will, allow the node to receive the header (address) information of new  nodes in the neighborhood using the same slot. 
One drawback of this approach is  its higher overhead  due frequent exchanges of control packets in case of  dynamic topology.

In \cite{D-Stabilization-Busch}  a  distributed and self-stabilizing TDMA-based MAC 
protocol is presented  which does not assume the global time reference. In this work, a randomized startup algorithm
with fault containment properties is used to perform the scheduling. 
Once the slots are determined, sensors
communicate among themselves to determine the period between successive slots or the frame size.
Each node divides the  time  into equal sized frames. However, the size of frame need not have to be same  across all the  nodes in the network. Also,  the frames do not need to be aligned at different nodes.
In this sense, the global time referencing is not required.
Furthermore, the algorithm proposed in  \cite{D-Stabilization-Busch} is also self-stabilizing, since it does not make any 
assumptions about the initial state. 
Table \ref{table:summary-distributed} provides the summary of surveyed distributed TDMA-scheduling algorithms 
as per the classification presented in section \ref{section:classification}.

\subsection{Cluster-based TDMA-Scheduling Algorithms}
The cluster-based TDMA MAC protocols commonly run in  rounds, where each round consists of a 
cluster set-up phase (TDMMA scheduling phase) and a steady-state  phase (data transmission phase). During the set-up phase, 
the nodes in the network are divided into groups called clusters. The clusters have special type of nodes called cluster 
heads (CH), which are responsible for TDMA-slot assignment among the nodes in the cluster. The set-up phase is followed
by a steady-state  phase, where the nodes can transmit their data using the slots allocated to them by their CHs.

One of the early work on clustering algorithm is by Heinzelman et.  al. \cite{clustered-Heinzelman}, in which they proposed a 
Low-Energy Adaptive Clustering Protocol (LEACH) for WSNs. LEACH is a distributed algorithm in which nodes make autonomous 
decisions without any centralized control. The goal is to maintain a constant number of clusters during each round and 
evenly distribute the load among all the node runs out of energy before others. The protocol assumes that every node in 
the network can reach the sink node with enough power.

The PACT protocol (Power Aware Clustered TDMA) \cite{clustered-Pei} proposed by Pei et. al. is one of the first TDMA MAC protocol 
for large sensor networks that used passive clustering in order to take advantage of a dense topology to prolong both 
battery and network lifetime. To improve the lifetime of the network, the PACT performs re-clustering of the network in a distributed manner considering the remaining battery energy level of the nodes. 
This is unlike LEACH algorithm,  in which the clustering status of 
a node is determined by the global knowledge of average number of clusters.
After clustering, a subset of cluster heads and certain 
gateway nodes are then selected which are responsible for traffic between neighboring clusters, have priority in allocating 
time slots.

In  \cite{clustered-Li} Li et. el. proposed a  MAC protocol called BMA for   intra-cluster even-based communication in large-scale cluster-based WSNs.  Similar to LEACH protocol the BMA protocol is divided into rounds and used the same algorithm for luster formation. After completing the cluster formation the system goes through with a series of phases with each phase consisting of 
three different periods namely contention period, a data transmission period and an idle period. The duration data transmission period is kept variable as every source node may not have the data to send all the time. The purpose of the contention period is to decide whether a node wants to transmit the data in the coming data transmission period or not.

\begin{landscape}
\def\baselinestretch{1}
\footnotesize
\tablefirsthead{
  \hline
   \textbf{Ref.} & \multicolumn{4}{|c|}{\textbf{Objective}} & \multicolumn{5}{|c|}{\textbf{Assumptions}} & 
   \multicolumn{3}{|c|}{\textbf{Design Methodology}} \\
  \hline
        & \textbf{MAC-Performance}  &  \multicolumn{3}{|c|}{\textbf{Schedule- Characteristics}} &  \multicolumn{2}{|c|}{\textbf{Application}} &  
        \multicolumn{2}{|c|}{\textbf{Topology}} &   
\textbf{Inter. Model} &  \textbf{Problem Formulation} & \textbf{Frequency of Resche.}  & \textbf{Type of Algo.}   \\
   \hline
     &  & \textbf{Slots/ Node} & \textbf{Conflict Free} & \textbf{Optim- ality} &  \textbf{Traffic Pattern} & 
     \textbf{Comm. Pattern} & \textbf{Logical Topo.} & \textbf{Physical} &  
   & & & \\
   \hline }

\tablehead{
 \hline
   \textbf{Ref.} & \multicolumn{4}{|c|}{\textbf{Objective}} & \multicolumn{5}{|c|}{\textbf{Assumptions}} & 
   \multicolumn{3}{|c|}{\textbf{Design Methodology}} \\
  \hline
        & \textbf{MAC-Performance}  &  \multicolumn{3}{|c|}{\textbf{Schedule- Characteristics}} &  \multicolumn{2}{|c|}{\textbf{Application}} &  
        \multicolumn{2}{|c|}{\textbf{Topology}} &   
\textbf{Inter. Model} &  \textbf{Problem Formulation} & \textbf{Frequency of Resche.}  & \textbf{Type of Algo.}   \\
   \hline
     &  & \textbf{Slots/ Node} & \textbf{Conflict Free} & \textbf{Optim- ality} &  \textbf{Traffic Pattern} & 
     \textbf{Comm. Pattern} & \textbf{Logical Topo.} & \textbf{Physical} &  
   & & & \\
   \hline
      }
   
\tabletail{%

\hline

\multicolumn{13}{|r|}{\small\sl continued on next page}\\

\hline}

\tablelasttail{\hline}

 \topcaption{Summary of Distributed TDMA-scheduling algorithms in WSNs}
\begin{supertabular}{|p{1.2cm}|p{1.5cm}|p{.9cm}|p{1.2cm}|p{1.7cm}|p{1.7cm}|p{1.3cm}|p{1.1cm}|p{1.1cm}|p{1.7cm}|p{2.0cm}|p{2.2cm}|p{1.1cm}|}
\label{table:summary-distributed}
& Maximize Throughput &   \multicolumn{11}{|c|}{\textbf{}} \\
\hline
\cite{D-MSL-Rhee,D-MSL-Kevin} & MSL & single & complete & sub- optimal & Periodic & Broadcast & 
Graph & Random & 2-hop  &  Conflict Graph & Multiple Frames & Det.  \\
\cite{D-MSL-Ranjeet} & MSL  & single & complete & sub- optimal & Periodic & Broadcast & Graph & Random 
 & 2-hop  &  Conflict Graph  & Multiple Frames & Det. \\ 
 \cite{D-MSL-Rajiv,D-MSL-Ephremides} & MTS & multiple & complete & sub- optimal & Periodic 
 & Broadcast & Graph & Random &  protocol (Tx Range) &  Conflict Graph & Multiple Frames & Det.  \\
\cite{D-dynamic-Cidon} & DS  & single & complete & feasible & variable load & Unicast/ Multicast & 
Graph & Random &  2-hop  & Conflict Graph & Per Frame & Det.   \\
\cite{D-dynamic-zhu} & DS & single & complete & feasible  & variable load & Broadcast & Graph & Random 
&  2-hop  & Conflict Graph & Per Frame & Det.  \\
\cite{D-dynamic-Chipara} & DS  & single & partial & feasible & variable load & Unicast & Tree & Random 
& 2-hop & Conflict Graph & Multiple Frames & Random  \\
\cite{D-dynamic-Salonidis} &  DS  & single & partial & feasible & variable load & Unicast & Tree & Random  
& 2-hop  & Conflict Graph & Per Frame & Random   \\
\hline \hline 
 & MO &   \multicolumn{11}{|c|}{\textbf{}} \\
 \hline
 \cite{D-MSL-Wang} &  MO  & single & complete & feasible & Periodic & Broadcast & Graph & Random & protocol (Tx 
 Range) & 
 Conflict Graph & Multiple Frames & Det.  \\
\cite{D-Overhead-Rozovsky,D-Overhead-Bao,D-Overhead-Rajendran} &  MO  & multiple & complete & feasible 
& variable load &  Broadcast & Graph & Random & protocol (Inter. Range) & Conflict graph (Use of PRNG) & Per Slot & Random    \\
 \cite{D-Overhead-Chih-Kuang} &  MO  & single & partial & feasible & periodic & Broadcast & Graph 
 & Random (logical grid) &  protocol (Tx. Range) & Latin Square (Use of position info) & Multiple Frames & Det.   \\
\hline \hline
 & Self Stabilization  &   \multicolumn{11}{|c|}{\textbf{}} \\
 \hline
 \cite{D-Stabilization-Ammar,D-Stabilization-Ali} & SS  & multiple & partial & feasible & periodic 
 &  broadcast & Graph & Random & 2-hop & Conflict Graph & Multiple Frames & Det.   \\
 \cite{D-Stabilization-Busch}  & SS  & single & complete & feasible & periodic &  broadcast & Graph & Random  
 & 2-hop &  Conflict Graph & Per Slot & Det.    \\
\end{supertabular}

\vspace{5pt}

\noindent MSL: Minimizing Schedule Length
\end{landscape}

\normalsize

If a node has a data to transmit, then it send the a 1-bit message in its allocated slot during the contention period. In this sense, the contention period also follows the TDMA based channel access. At the end of the contention period the cluster would have all the information about the data transmission requirements of nodes belonging to the  that cluster. Based on this information the cluster head decides the length of data transmission period and broadcasts the TDMA-schedule to all the nodes in the cluster who had request a transmission slot.
After receiving the data from source nodes the cluster head then aggregates and forwards the data to the base station.

Tavli et. al. in  \cite{clustered-Tavli} propose a clustering scheme MH-TRACE, which is not based on connectivity
information.  In MH-TRACE, cluster creation  and  maintenance does not require explicit exchange of 
connectivity information among neighboring nodes.  Instead,  the clustering algorithm  continuously monitors   the level of interference and take the actions accordingly to minimize the inter-cluster interference.
This technique incurs less  overhead as other clustering approaches as they involve the transmission of connectivity information among the nodes.

Biaz et. al. in \cite{clustered-Biaz} propose  a cluster-based MAC protocol to resolve the contention for forwarded traffic. The protocol is based on the assumption that in WSNs the volume of forwarded traffic is much more than the originated traffic.
There are two types of slots in a frame. One type of slots are used by are dedicated for cluster heads and use TDMA-based channel access mechanism. The other types of slots are for non-cluster nodes and  here the contention-based  scheme is used to resolve the conflict. The number of TDMA-slots in each time is is kept much larger than the contention-slots assuming that the cluster heads are required to transmit more data than the non-cluster nodes. The idea of reserving the bandwidth for cluster heads reduced the contention caused by the inter-cluster communication.

 Haigang et. al. in \cite{clustered-Haigang} propose an interference free TDMA-scheduling algorithm for cluster base WSNs. 
The time is divided into superframes, where each superframe consist of multiple TDMA frames. Every TDMA frame is further divided into multiple time-slots.  Different  frames in a superframe are alloted to the neighboring clusters to avoid inter cluster interference.    Further,  different time slots of a frame are given to the nodes belonging the cluster to which the frame has been alloted. This avoids 
intra-cluster interference. In this way,  the  neighboring cluster heads collect the data from their sensor nodes  during different TDMA frames to avoid inter-cluster interference.

\section{Conclusions}
\label{section:LITERATURE_Conclusion}
In this paper, we have presented a classification framework to understand the various aspects of existing TDMA scheduling 
algorithms in WSNs. The framework is based on the classification of the features of TDMA-scheduling algorithms  
into three categories, viz. objective, assumptions, and design methodology. This type of classification  is especially useful 
to understand design space (problem space and the solution space) of a TDMA-scheduling algorithm in WSNs
since many scheduling algorithms available in the literature does not state several aspects of the scheduling 
problem explicitly. Additionally, the framework is  useful to compare these protocols in a qualitative manner. 
Furthermore, we discussed various TDMA-scheduling  protocols, and provided a summary of their characteristics
(Table \ref{table:summary-centralized}, \ref{table:summary-distributed}) based on proposed classification framework.


\begin{thebibliography}{10}
\providecommand{\url}[1]{{#1}}
\providecommand{\urlprefix}{URL }
\expandafter\ifx\csname urlstyle\endcsname\relax
  \providecommand{\doi}[1]{DOI~\discretionary{}{}{}#1}\else
  \providecommand{\doi}{DOI~\discretionary{}{}{}\begingroup
  \urlstyle{rm}\Url}\fi

\bibitem{Latin}
{Latin square}.
\newblock http://en.wikipedia.org/wiki/Latin\_square.
\newblock Accessed 30-Dec-2014

\bibitem{Vizing}
{Vizing's Theorem}.
\newblock http://en.wikipedia.org/wiki/Graph \_coloring.
\newblock Accessed 30-Dec-2014

\bibitem{D-Stabilization-Ali}
Ali, F.N., Appani, P.K., Hammond, J.L., Mehta, V.V., Noneaker, D., Russell, H.:
  Distributed and adaptive tdma algorithms for multiple-hop mobile networks.
\newblock In: Military Communications Conference, MILCOM 2002., vol.~1, pp.
  546--551. IEEE (2002)

\bibitem{D-Stabilization-Ammar}
Ammar, M., Stevens, D.: A distributed tdma rescheduling procedure for mobile
  packet radio networks.
\newblock In: IEEE International Conference on Communications, 1991. ICC '91,
  pp. 1609--1613 vol.3 (1991)

\bibitem{C-MSL-Annamalai}
Annamalai, V., Gupta, S., Schwiebert, L.: On tree-based convergecasting in
  wireless sensor networks.
\newblock In: Wireless Communications and Networking, 2003. WCNC 2003. 2003
  IEEE, vol.~3, pp. 1942--1947 (2003)

\bibitem{D-Overhead-Bao}
Bao, L., Garcia-Luna-Aceves, J.J.: A new approach to channel access scheduling
  for ad hoc networks.
\newblock In: the 7th Annual International Conference on Mobile Computing and
  Networking, MobiCom '01, pp. 210--221. ACM, New York, NY, USA (2001)

\bibitem{DTSS}
Bhatia, A., Hansdah, R.C.: A distributed tdma slot scheduling algorithm for
  spatially correlated contention in wsns.
\newblock In: 2013 27th International Conference on Advanced Information
  Networking and Applications Workshops, pp. 377--384 (2013).
\newblock \doi{10.1109/WAINA.2013.23}

\bibitem{RDTDMA}
Bhatia, A., Hansdah, R.C.: Rd-tdma: A randomized distributed tdma scheduling
  for correlated contention in wsns.
\newblock In: 2014 28th International Conference on Advanced Information
  Networking and Applications Workshops, pp. 378--384 (2014)

\bibitem{DSLR}
Bhatia, A., Hansdah, R.C.: Dslr: A distributed schedule length reduction
  algorithm for wsns.
\newblock In: 2015 IEEE International Parallel and Distributed Processing
  Symposium, pp. 365--374 (2015)

\bibitem{clustered-Biaz}
Biaz, S., Barowski, Y.D.: Gangs: an energy efficient mac protocol for sensor
  networks.
\newblock In: Proceedings of the 42nd annual Southeast regional conference, pp.
  82--87. ACM (2004)

\bibitem{D-MSL-Kevin}
Bryan, K.L., Ren, T., Henry, T., Fay-wolfe, V.: Towards optimal tdma frame size
  in wireless sensor networks, technical report, university of rhode island
  (2007)

\bibitem{D-Stabilization-Busch}
Busch, C., Magdon-Ismail, M., Sivrikaya, F., Yener, B.: Contention-free mac
  protocols for wireless sensor networks.
\newblock In: R.~Guerraoui (ed.) Distributed Computing, \emph{Lecture Notes in
  Computer Science}, vol. 3274, pp. 245--259. Springer Berlin Heidelberg (2004)

\bibitem{C-MF-Chatterjea}
Chatterjea, S., van Hoesel, L., Havinga, P.: Ai-lmac: an adaptive,
  information-centric and lightweight mac protocol for wireless sensor
  networks.
\newblock In: Intelligent Sensors, Sensor Networks and Information Processing
  Conference, 2004. Proceedings of the 2004, pp. 381--388 (2004).
\newblock \doi{10.1109/ISSNIP.2004.1417492}

\bibitem{D-dynamic-Chipara}
Chipara, O., Lu, C., Stankovic, J., Roman, G.: Dynamic conflict-free
  transmission scheduling for sensor network queries.
\newblock IEEE Transactions onMobile Computing \textbf{10}(5), 734--748 (2011)

\bibitem{C-MSL-Choi}
Choi, H., Wang, J., Hughes, E.: Scheduling for information gathering on sensor
  network.
\newblock Wireless Networks \textbf{15}(1), 127--140 (2009).
\newblock \doi{10.1007/s11276-007-0050-9}.
\newblock \urlprefix\url{http://dx.doi.org/10.1007/s11276-007-0050-9}

\bibitem{D-dynamic-Cidon}
Cidon, I., Sidi, M.: Distributed assignment algorithms for multihop packet
  radio networks.
\newblock IEEE Transactions on Computers \textbf{38}(10), 1353--1361 (1989)

\bibitem{C-ML-Cui}
Cui, S., Madan, R., Goldsmith, A., Lall, S.: Energy-delay tradeoffs for data
  collection in tdma-based sensor networks.
\newblock In: Communications, 2005. ICC 2005. 2005 IEEE International
  Conference on, vol.~5, pp. 3278--3284 Vol. 5 (2005)

\bibitem{D-MSL-Ephremides}
Ephremides, A., Truong, T.: Scheduling broadcasts in multihop radio networks.
\newblock IEEE Transactions on Communications \textbf{38}(4), 456--460 (1990)

\bibitem{C-MSL-Ergen}
Ergen, S.C., Varaiya, P.: Tdma scheduling algorithms for wireless sensor
  networks.
\newblock Wireless Networks \textbf{16}(4), 985--997 (2010)

\bibitem{C-MSL-Florens1}
Florens, C., Franceschetti, M., McEliece, R.: Lower bounds on data collection
  time in sensory networks.
\newblock IEEE Journal onSelected Areas in Communications \textbf{22}(6),
  1110--1120 (2004)

\bibitem{C-MSL-Florens2}
Florens, C., McEliece, R.: Scheduling algorithms for wireless ad-hoc sensor
  networks.
\newblock In: IEEE Global Telecommunications Conference, 2002. GLOBECOM '02.,
  vol.~1, pp. 6--10 vol.1 (2002)

\bibitem{C-MSL-Florens3}
Florens, C., McEliece, R.: Packets distribution algorithms for sensor networks.
\newblock In: INFOCOM 2003. Twenty-Second Annual Joint Conference of the IEEE
  Computer and Communications., vol.~2, pp. 1063--1072 vol.2 (2003)

\bibitem{survey3-Gabale}
Gabale, V., Raman, B., Dutta, P., Kalyanraman, S., Raman, B., Dutta, P.,
  Kalyanraman, S.: A classification framework for scheduling algorithms in
  wireless mesh networks.
\newblock Communications Surveys Tutorials, IEEE \textbf{15}(1), 199--222
  (2013)

\bibitem{c-ML-gandham}
Gandham, S., Zhang, Y., Huang, Q.: Distributed time-optimal scheduling for
  convergecast in wireless sensor networks.
\newblock Computer Networks \textbf{52}(3), 610--629 (2008)

\bibitem{Gronkvist}
Gr\"{o}nkvist, J., Hansson, A.: Comparison between graph-based and
  interference-based stdma scheduling.
\newblock In: Proceedings of the 2Nd ACM International Symposium on Mobile Ad
  Hoc Networking \&Amp; Computing, MobiHoc '01, pp. 255--258. ACM, New York,
  NY, USA (2001).
\newblock \doi{10.1145/501449.501453}.
\newblock \urlprefix\url{http://doi.acm.org/10.1145/501449.501453}

\bibitem{Gupta}
Gupta, P., Kumar, P.: The capacity of wireless networks.
\newblock Information Theory, IEEE Transactions on \textbf{46}(2), 388--404
  (2000).
\newblock \doi{10.1109/18.825799}

\bibitem{clustered-Haigang}
Haigang, G., Ming, L., Xiaomin, W., Lijun, C., Li, X.: An interference free
  cluster-based tdma protocol for wireless sensor networks.
\newblock In: X.~Cheng, W.~Li, T.~Znati (eds.) Wireless Algorithms, Systems,
  and Applications, \emph{Lecture Notes in Computer Science}, vol. 4138, pp.
  217--227. Springer Berlin Heidelberg (2006)

\bibitem{clustered-Heinzelman}
Heinzelman, W., Chandrakasan, A., Balakrishnan, H.: Energy-efficient
  communication protocol for wireless microsensor networks.
\newblock In: System Sciences, 2000. Proceedings of the 33rd Annual Hawaii
  International Conference on, pp. 10 pp. vol.2-- (2000).
\newblock \doi{10.1109/HICSS.2000.926982}

\bibitem{LMAC}
van {Hoesel}, L., {Havinga}, P.: A lightweight medium access protocol (lmac)
  for wireless sensor networks: Reducing preamble transmissions and transceiver
  state switches.
\newblock In: 1st International Workshop on Networked Sensing Systems, INSS
  2004, pp. 205--208. Society of Instrument and Control Engineers (SICE),
  Tokio, Japan (2004).
\newblock \urlprefix\url{http://doc.utwente.nl/64756/}

\bibitem{survey3-Incel}
Incel, O., Ghosh, A., Krishnamachari, B.: Scheduling algorithms for tree-based
  data collection in wireless sensor networks.
\newblock In: S.~Nikoletseas, J.D. Rolim (eds.) Theoretical Aspects of
  Distributed Computing in Sensor Networks, Monographs in Theoretical Computer
  Science. An EATCS Series, pp. 407--445. Springer Berlin Heidelberg (2011)

\bibitem{c-ME-Kalpakis}
Kalpakis, K., Dasgupta, K., Namjoshi, P.: Efficient algorithms for maximum
  lifetime data gathering and aggregation in wireless sensor networks.
\newblock Computer Networks \textbf{42}(6), 697 -- 716 (2003).
\newblock \doi{http://dx.doi.org/10.1016/S1389-1286(03)00212-3}.


\bibitem{clustered-Li}
Li, J., Lazarou, G.Y.: A bit-map-assisted energy-efficient mac scheme for
  wireless sensor networks.
\newblock In: Proceedings of the 3rd international symposium on Information
  processing in sensor networks, pp. 55--60. ACM (2004)

\bibitem{D-Overhead-Chih-Kuang}
Lin, C.K., Zadorozhny, V., Krishnamurthy, P., Park, H.H., Lee, C.G.: A
  distributed and scalable time slot allocation protocol for wireless sensor
  networks.
\newblock IEEE Transactions on Mobile Computing \textbf{10}(4), 505--518 (2011)

\bibitem{C-ML-Gang}
Lu, G., Krishnamachari, B.: Minimum latency joint scheduling and routing in
  wireless sensor networks.
\newblock Ad Hoc Networks \textbf{5}(6), 832 -- 843 (2007).
\newblock \doi{http://dx.doi.org/10.1016/j.adhoc.2007.03.002}.


\bibitem{C-MSL-Malhotra}
Malhotra, B., Nikolaidis, I., Nascimento, M.A.: Aggregation convergecast
  scheduling in wireless sensor networks.
\newblock Wirel. Netw. \textbf{17}(2), 319--335 (2011).
\newblock \doi{10.1007/s11276-010-0282-y}.
\newblock \urlprefix\url{http://dx.doi.org/10.1007/s11276-010-0282-y}

\bibitem{c-ME-Mao}
Mao, J., Wu, Z., Wu, X.: A \{TDMA\} scheduling scheme for many-to-one
  communications in wireless sensor networks.
\newblock Computer Communications \textbf{30}(4), 863 -- 872 (2007).
\newblock \doi{http://dx.doi.org/10.1016/j.comcom.2006.10.006}.
\newblock Nature-Inspired Distributed Computing

\bibitem{Moscibroda}
Moscibroda, T.: The worst-case capacity of wireless sensor networks.
\newblock In: Information Processing in Sensor Networks, 2007. IPSN 2007. 6th
  International Symposium on, pp. 1--10 (2007).
\newblock \doi{10.1109/IPSN.2007.4379659}

\bibitem{C-ML-Pan}
Pan, M.S., Tseng, Y.C.: Quick convergecast in zigbee beacon-enabled tree-based
  wireless sensor networks.
\newblock Computer Communications \textbf{31}(5), 999 -- 1011 (2008).
\newblock \doi{http://dx.doi.org/10.1016/j.comcom.2007.12.015}.
\newblock Mobility Management and Wireless Access

\bibitem{D-MSL-Ranjeet}
Patro, R., Mohan, B.: Mobile agent based tdma slot assignment algorithm for
  wireless sensor networks.
\newblock In: International Conference on Information Technology: Coding and
  Computing, 2005. ITCC 2005., vol.~2, pp. 663--667 Vol. 2 (2005)

\bibitem{clustered-Pei}
Pei, G., Chien, C.: Low power tdma in large wireless sensor networks.
\newblock In: Military Communications Conference, 2001. MILCOM 2001.
  Communications for Network-Centric Operations: Creating the Information
  Force. IEEE, vol.~1, pp. 347--351 vol.1 (2001).
\newblock \doi{10.1109/MILCOM.2001.985817}

\bibitem{D-Overhead-Rajendran}
Rajendran, V., Obraczka, K., Garcia-Luna-Aceves, J.J.: Energy-efficient,
  collision-free medium access control for wireless sensor networks.
\newblock Wireless Networks \textbf{12}(1), 63--78 (2006).
\newblock \doi{10.1007/s11276-006-6151-z}.
\newblock \urlprefix\url{http://dx.doi.org/10.1007/s11276-006-6151-z}

\bibitem{C-MSL-Ramanathan}
Ramanathan, S.: A unified framework and algorithm for (t/f/c)dma channel
  assignment in wireless networks.
\newblock In: INFOCOM '97. Sixteenth Annual Joint Conference of the IEEE
  Computer and Communications Societies. Driving the Information Revolution.,
  vol.~2, pp. 900--907 vol.2 (1997)

\bibitem{D-MSL-Rajiv}
Ramaswami, R., Parhi, K.: Distributed scheduling of broadcasts in a radio
  network.
\newblock In: Proceedings of the Eighth Annual Joint Conference of the IEEE
  Computer and Communications Societies. Technology: Emerging or Converging,
  IEEE INFOCOM '89., pp. 497--504 vol.2 (1989)

\bibitem{C-ML-Revah}
Revah, Y., Segal, M.: Improved lower bounds for data-gathering time in sensor
  networks.
\newblock In: Third International Conference on Networking and Services, 2007.
  ICNS., pp. 76--76 (2007)

\bibitem{D-MSL-Rhee-ZMAC}
Rhee, I., Warrier, A., Aia, M., Min, J., Sichitiu, M.: Z-mac: A hybrid mac for
  wireless sensor networks.
\newblock Networking, IEEE/ACM Transactions on \textbf{16}(3), 511--524 (2008).
\newblock \doi{10.1109/TNET.2007.900704}

\bibitem{D-MSL-Rhee}
Rhee, I., Warrier, A., Min, J., Xu, L.: Drand: distributed randomized tdma
  scheduling for wireless ad-hoc networks.
\newblock In: Proceedings of the 7th ACM international symposium on Mobile ad
  hoc networking and computing, MobiHoc '06, pp. 190--201. ACM (2006)

\bibitem{D-Overhead-Rozovsky}
Rozovsky, R., Kumar, P.R.: Seedex: A mac protocol for ad hoc networks.
\newblock In: 2Nd ACM International Symposium on Mobile Ad Hoc Networking
  \&Amp; Computing, MobiHoc '01, pp. 67--75. ACM, New York, NY, USA (2001).
\newblock \doi{10.1145/501426.501427}.
\newblock \urlprefix\url{http://doi.acm.org/10.1145/501426.501427}

\bibitem{D-dynamic-Salonidis}
Salonidis, T., Tassiulas, L.: Distributed dynamic scheduling for end-to-end
  rate guarantees in wireless ad hoc networks.
\newblock In: 6th ACM International Symposium on Mobile Ad Hoc Networking and
  Computing, MobiHoc '05, pp. 145--156. ACM, New York, NY, USA (2005).
\newblock \doi{10.1145/1062689.1062709}.
\newblock \urlprefix\url{http://doi.acm.org/10.1145/1062689.1062709}

\bibitem{C-MF-Sridharan}
Sridharan, A., Krishnamachari, B.: Max-min fair collision-free scheduling for
  wireless sensor networks.
\newblock In: Performance, Computing, and Communications, 2004 IEEE
  International Conference on, pp. 585--590 (2004).
\newblock \doi{10.1109/PCCC.2004.1395103}

\bibitem{clustered-Tavli}
Tavli, B., Heinzelman, W.: Mh-trace: multihop time reservation using adaptive
  control for energy efficiency.
\newblock Selected Areas in Communications, IEEE Journal on \textbf{22}(5),
  942--953 (2004).
\newblock \doi{10.1109/JSAC.2004.826932}

\bibitem{survey3-Wang}
Wang, L., Xiao, Y.: A survey of energy-efficient scheduling mechanisms in
  sensor networks.
\newblock Mobile Networks and Applications \textbf{11}(5), 723--740 (2006).
\newblock \doi{10.1007/s11036-006-7798-5}

\bibitem{D-MSL-Wang}
Wang, Y., Henning, I.: A deterministic distributed tdma scheduling algorithm
  for wireless sensor networks.
\newblock In: International Conference on Wireless Communications, Networking
  and Mobile Computing, 2007. WiCom 2007., pp. 2759 --2762 (2007)

\bibitem{D-dynamic-zhu}
Zhu, C., Corson, M.S.: A five-phase reservation protocol (fprp) for mobile ad
  hoc networks.
\newblock Wireless Networks \textbf{7}(4), 371--384 (2001).
\newblock \doi{10.1023/A:1016683928786}.
\newblock \urlprefix\url{http://dx.doi.org/10.1023/A:1016683928786}

\end{thebibliography}
\end{document}